\documentclass[twocolumn,prl,aps,preprintnumbers,notitlepage]{revtex4-1}
\usepackage[latin9]{inputenc}
\usepackage{amsmath}
\usepackage{amssymb}
\usepackage{graphicx}
\usepackage{esint}
\usepackage[scr=boondoxo, scrscaled=1.05]{mathalfa}
\allowdisplaybreaks

\makeatletter

\pdfpageheight\paperheight
\pdfpagewidth\paperwidth

\pdfoutput=1

\usepackage{epsfig}
\usepackage{graphics}

\usepackage{mciteplus}
\mciteErrorOnUnknownfalse

\makeatother

\begin{document}

\preprint{FERMILAB-PUB-18-178-A}

\title{Nonlocal Entanglement and Directional Correlations of  Primordial Perturbations on the Inflationary Horizon}
\author{Craig  Hogan}
\affiliation{University of Chicago and Fermilab}

\begin{abstract}
Models  are developed to estimate  properties of  relic cosmic perturbations with ``spooky''  nonlocal correlations on the inflationary horizon, analogous to those previously posited for  information on black hole event horizons.    Scalar curvature perturbations are estimated to emerge with a dimensionless power spectral density $\Delta_S^2\approx H t_P$, the product of inflationary expansion rate $H$ with Planck time $t_P$,  larger than  standard inflaton fluctuations.  Current measurements of the spectrum are used to derive  constraints on parameters of the effective potential in a slow-roll background.   It is shown that  spooky nonlocality can create statistically homogeneous and isotropic primordial curvature perturbations that are initially  directionally antisymmetric.  New statistical estimators are developed to study unique signatures   in CMB anisotropy and large scale galaxy surveys.
\end{abstract}

\maketitle

\section{Introduction} 

Cosmic perturbations on the largest scales are widely thought to come from microscopic quantum fluctuations on the horizon scale during  inflation.
This hypothesis is supported by a unique and precisely measured experimental  signature,  a power spectrum of primordial curvature perturbations on very large scales that is almost but not exactly scale-free\cite{Ade:2015xua,Ade:2015lrj,Array:2015xqh,Akrami:2018vks,Aghanim:2018eyx,Akrami:2018odb}.

To account for these data,  slow-roll inflation\cite{Kadota:2005hv,Baumann:2009ds,Kamionkowski:2015yta} posits  
a classical background universe   that expands nearly exponentially according to classical general relativity, driven by  the free energy density $V(\phi)$  of a nearly-uniform  scalar field  with a slowly time-varying classical expectation value, $\phi$. 
In this setting, the  quantum model that leads to    perturbations is adapted from high energy particle physics: curvature perturbations are produced by the gravitation of quantum field fluctuations when they freeze out on the inflationary horizon scale. 
Standard inflation models  assume that  quantum geometrical degrees of freedom behave like those of  quantum fields, and  that classical properties of space and time are well defined and determinate on all scales.

A new  hypothesis about the primordial quantum system  is explored here: {\it the entire inflationary horizon is assumed to be a single quantum object}.
Even properties of space and time that are universal to all classical metrics, such as a local inertial frame, are  allowed  to be nonlocal and indeterminate, so  perturbations can emerge with new kinds of ``spooky''  nonlocal  correlations that are classically impossible.   The standard model of inflation and  linear perturbation mode evolution is still assumed on all scales after they  exit the inflationary horizon.


Similar nonlocal  quantum coherence of horizon states  has  been invoked to resolve  information paradoxes in evaporating particle states that create back reaction on  black hole event horizons\cite{Hooft:2016cpw,Hooft:2016itl,Hooft2018}.

 In this model,  the origin of cosmic perturbations is not separate from the emergence of locality and of space-time itself from a quantum system.
Classical space-time, along with its local inertial frame and the local cosmic standard of rest, emerge together as a holistic process. On the inflationary horizon, geometrical quantum states are nonlocal and include new kinds of entanglement among all directions.
The emergent perturbations of  classical  invariant curvature  display  previously-neglected, nonlocally correlated noise.  Their  nonlocal, multidirectional  correlations  can have measurable physical effects on the  phases of relic perturbations.

Specific properties of spooky correlations  are estimated  here by adapting covariant  models  of locality, emergence, and entanglement previously developed to design and interpret  laboratory experiments\cite{Hogan:2015b,Hogan:2016,0264-9381-35-20-204001}, based on  Planck scale quantum states with  nonlocal correlations that extend everywhere on light cones or  spacelike causal diamond surfaces.
The relic curvature perturbations are estimated to exceed the standard, inflaton-generated perturbations by a significant factor.   The estimated emergent perturbation spectrum agrees with current measurements.   The model has fewer parameters than  standard  inflation models, since perturbations arise from  a quantum-geometrical effect that is not sensitive to properties of the matter fields.

Some still-untested predictions differ nontrivially from  standard  inflation.  Signatures of spooky primordial correlations can survive in  cosmic density perturbations today, in particular, a new kind of scale-free directional antisymmetry that violates locality at inflation.    Specific  model-independent statistical tests can  distinguish spooky correlations from  standard perturbations on scales still in the linear regime.  
It is suggested below that they  might already be  detected in CMB anistropy, and if so, that  it may be possible to detect them with new kinds of  measurements  with  large scale galaxy surveys.

\section{Spooky  inflation}

\subsection{Homogeneous classical inflation}

  A  standard inflation model\cite{Kadota:2005hv,Baumann:2009ds,Kamionkowski:2015yta} is   assumed throughout this paper  for the classical background cosmology.  The model of mass-energy is  a spatially uniform classical (that is, unquantized) inflaton field, with dimension of mass and vacuum expectation value $\phi(t)$, where $t$ is a standard FRW time coordinate.
  In standard notation where $\hbar= c =1$, the expansion rate $H$ and cosmic scale factor $a$  evolve according to classical general relativity and thermodynamics,
\begin{equation}\label{hubble}
H^2(t)\equiv (\dot a/a)^2= (8\pi G/3) (V(\phi)+\dot \phi^2/2),
\end{equation}
where the evolution of the  inflaton depends on the potential $V(\phi)$  via
\begin{equation}\label{roll}
\ddot \phi + 3H\dot \phi + V' = 0,
\end{equation}
and  $V'\equiv dV/d\phi$.
During  slow roll inflation,  the evolution of $\phi$ approximately obeys
\begin{equation}\label{slowroll}
3H\dot \phi \approx  - V',
\end{equation}
which produces  
a nearly-exponential expansion.
About 60 e-foldings in $a$ after the currently observable volume of the universe matches the scale $c/H$ of the inflationary horizon, inflation ends,  and subsequently  ``reheats'' with the conversion of $\phi$ to other forms of matter.

Quantum fluctuations of the inflaton, although they are presumably still present for a physical inflaton field,  are neglected here; as shown below,  their gravitational effect is  smaller than  the spooky geometrical perturbations. As usual, perturbations of  wavenumber $k$   freeze in at the cosmic scale factor $a(k)$ when  $k= a(k) H(k) /c $. 

The background evolution at late times is assumed to be the standard concordance $\Lambda$CDM model.
This standard background solution provides the  global
definition of surfaces of unperturbed cosmic time on comoving world lines, corresponding to surfaces where $\phi$  is constant.

\subsection{Spooky correlations in emergent gravity}

At the most basic level, quantum mechanics is a theory of correlations that does not  assume any particular projection onto space and time. It is possible, 
as envisioned in  relational (or ``emergent'') quantum gravity, that locality--- the relationship that differentiates  space-time positions or events---   emerges  as an approximate observable in a quantum system\cite{Rovelli2004,0264-9381-28-15-153002,Banks:2018aed}. In general, relational quantum gravitational  degrees of freedom and correlations  differ from those of fields.  
They can  produce quantum fluctuations associated with nonlocal correlations of positional relationships on all scales.

\subsubsection{Precedents for nonlocal correlations of quantum geometry}

Theoretical studies, especially of systems with horizons, have long hinted that  space-time relationships are encoded in  entanglement  information,  analogous to  spooky macroscopic correlations of entangled particle states. If space and time emerge from a quantum system,  a new  kind of nonlocal correlation on all scales is needed to account for
finite and holographic  gravitational information in black holes\cite{Bekenstein1973,Hawking1975,Wald:1999vt}; its generalization to a ``holographic principle'' in any space-time\cite{tHooft:1993dmi,Susskind1995,Bousso:2002ju};   consistent evolution of matter fields and information flow in the presence of black hole horizons\cite{Hooft:2016cpw,Hooft:2016itl,Hooft2018} without information paradoxes\cite{0034-4885-80-9-092002};  the absence of  field states more massive  than black holes in a volume of any size\cite{CohenKaplanNelson1999}; and   holographic correlations  in anti-de Sitter space\cite{Ryu:2006bv,Ryu:2006ef,Solodukhin:2011gn,Natsuume:2014sfa}.   

 These results suggest that information in quantum geometrical degrees of freedom is less localized, and more universally entangled, than that in particles and fields, even though it is governed by a much smaller dimensional scale, the Planck time $t_P\equiv \sqrt{\hbar G/ c^5} = 5.4\times 10^{-44}$sec. 
 It is even possible to derive general relativity thermodynamically,  as a statistical  theory or equation of state\cite{Jacobson1995,Verlinde2011,Padmanabhan:2013nxa,Jacobson:2015hqa}, 
where the basic elements are invariant null surfaces, such as horizons,  light cones, and causal diamonds\cite{tHooft:1993dmi,Susskind1995,Bousso:2002ju}.
As elegantly  prefigured by Wheeler\cite{10.2307/3301037},  ``\dots in the gravitational theory we should be able in principle to dispense with the concepts of space and time and take as the basis of our description of nature the elementary concepts of world line and light cone.''

\subsubsection{Physical effects of exotic  correlations}

The previous considerations are all of a general, abstract nature. 
No consensus exists about  concrete physical  effects of  exotic holographic geometrical correlations on large scales, and no experimental departure from classical space-time has been convincingly demonstrated.  
Even so, there are  theoretical and experimental constraints on the specific form exotic correlations  can take.

In standard quantum mechanics,  ``spooky action at a distance'' refers to nonlocal quantum 
correlations of entangled particle states that
 extend indefinitely in the future  of an event where a  state is prepared\cite{RevModPhys.71.S288,Handsteiner:2016ulx}.
For example, in
 positron emission tomography, 
the space-time position of an annihilation event can be reconstructed,  in principle with diffraction-limited fidelity,  from a macroscopic  correlation in arrival times  and positions of a pair of entangled  photons  traveling in opposite directions anywhere on its future light cone. 

Entangled particle pairs act as sources for superpositions of gravitational states, so geometry itself must also  have spooky nonlocal correlations on light cones.  Unlike the particle example, geometrical states are universal: they must entangle with all forms of matter and energy on a light cone in all directions, not just a single pair of particles.  Their correlations describe the relationship of  the  local inertial frame of a world line to  the rest of the universe\cite{Hogan:2015b,Hogan:2016,0264-9381-35-20-204001}.

Exotic correlations of geometry must  exist in flat space-time as well as black holes, so they should affect  states of light in the laboratory.  
They could have escaped experimental detection    
because the estimated correlation scale is very small---  comparable not to the Planck length, but to the diffraction width of a Planck bandwidth wave function\cite{Hogan:2010zs}. Even so,  they
 might be measurable with new kinds of experiments\cite{Holo:Instrument,2013PhRvL.110u3601R,PhysRevA.92.053821,Pradyumna:2018xbx}.  Indeed, experimental constraints on  symmetries of Planck scale tensor-like holographic correlations\cite{Holo:PRL,holoshear} are used below to constrain predictions of tensor modes in spooky primordial perturbations.

 \begin{figure}
\begin{centering}
\includegraphics[width=\linewidth]{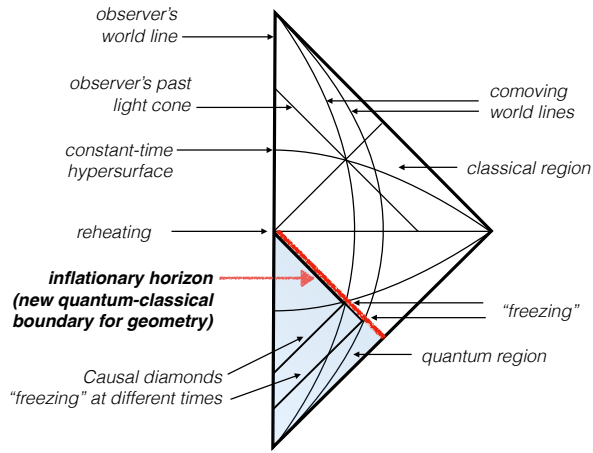}
\par\end{centering}
\protect\caption{ Penrose diagram of the  standard inflationary $\Lambda$CDM universe. Constant time and space surfaces are shown in comoving coordinates.  In the spooky scenario, the quantum-classical boundary for geometry lies on an observer's inflationary horizon, the null surface represented by the upper boundary of the shaded region.  Entanglement  on the horizon creates new, delocalized  spooky correlations of perturbations among different   spatial directions at reheating. \label{penrose}}
\end{figure}

 \begin{figure}
\begin{centering}
\includegraphics[width=\linewidth]{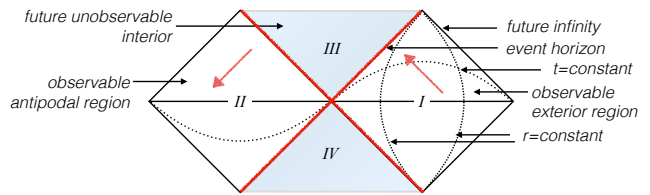}
\par\end{centering}
\protect\caption{ Maximally extended Penrose diagram of an eternal Schwarzschild black hole, adapted from ref. \cite{Hooft:2016cpw}. Surfaces  of constant time $t$ and radius $r$ are shown for Schwarzschild coordinates that approach proper coordinates  for a distant external observer. Entanglement on the horizon creates globally delocalized correlations between positions and momenta of incoming and outgoing particle states, time-antisymmetric between antipodal regions $I$ and $II$, as indicated by bold arrows. In spooky inflation, similar antipodal time antisymmetry on the inflationary horizon can lead to directionally antisymmetric scalar curvature perturbations. \label{blackhole}}
\end{figure}

\subsubsection{Quantum models of  inflationary fluctuations}

Calculations of perturbations in standard  inflation use a model quantum system  based on local quantum field theories originally developed for high energy particle interactions  (e.g., ref. \cite{STAROBINSKY198099}).
The quantized system is the amplitude of a field in space-time, often described as a superposition of modes,  each one of which approximates a quantized  harmonic oscillator.
The standard model system  violates locality in a particular way: each mode has built-in spacelike correlations, the classical spatial structure of a plane wave with a certain wavenumber.  In standard calculations, this is often expressed mathematically by writing the initial state as a field vacuum state in comoving  coordinates.

This model quantum system is not adequate to include all the correlations that could occur among space-time degrees of freedom.
The background space-time is classical, meaning that positional relationships are described by commuting quantities.
The same classical locality is assigned to  quantum fields and their gravitational effects: although the amplitudes are quantized, the comoving field modes have a determinate, globally-defined spatial structure that is shared by the relic metric perturbations.

The  new  hypothesis here is   that during inflation, locality does not apply down to the Planck scale, only down to the horizon scale.  
Before then, space-time is not constrained to be a classical differentiable manifold.  Primordial correlations can violate locality in new ways:
 relative positions and proper times of comoving world lines  emerge with spatially nonlocal correlations at reheating when they become classical.  
 
 The new quantum-classical boundary for perturbations (Fig. (\ref{penrose}) is defined by the classical causal structure around each observer. The ``outgoing'' states  are represented by world lines when they pass through the horizon. In the inflationary context, freezing of perturbations is the equivalent of collapse or measurement in the laboratory, and outgoing states ({\it i.e.}, the positions of world lines) are  entangled with each other everywhere on the horizon.  In this respect, 
the nonlocal entanglement  of perturbations emerging from  inflationary horizon states resembles global entanglement of incoming and outgoing particle states emerging from black hole  horizons, introduced to solve information paradoxes\cite{Hooft:2016cpw,Hooft:2016itl,Hooft2018}.  An eternal  black hole horizon (Fig. \ref{blackhole}) creates directionally  antisymmetric correlations among particle states from  quantum back reaction; in spooky inflation, the  horizon  creates directionally antisymmetric  curvature perturbations.

In standard inflation, the  quantum-classical boundary is the same for all observers:
each plane wave mode of fixed comoving size and direction ``freezes'' everywhere at the same comoving time,  with  spatial relationships among world lines determined by a determinate classical background.  
Here, the emergent space-time hypothesis  implies an observer-dependent boundary of the horizon and the quantum region.
For any two world lines, their classical positions only freeze in much later, at a time specific to  their locations and separation direction.
This indeterminacy allows a nonlocal spacelike entanglement among different directions that cannot occur for standard field modes.


The main goal of this paper is to show that  it is possible to incorporate these new features into a consistent model for  emergent classical perturbations. The models developed here allow sharper predictions than previous generic estimates of holographic discreteness effects on inflation\cite{Hogan:2002xs,Hogan:2003mq}.
As shown below,  spooky quantum fluctuations  project onto observable modes in a way that introduces larger perturbations than usual, and introduces previously-forbidden antisymmetric correlations.

Ultimately, a full theory will require a new quantum model  that can include interference in three directions and  a  model of freezing that can account for the classical causal structure and local inertial frame that emerges for each world-line.
A   model  of quantum-geometrical  states cannot be based on a standard correspondence principle, since they represent new unknown quantum degrees of freedom of emergent space and time.  Here, simple models are  developed based on constraints from  matching to classical symmetries--- first for causal diamonds in flat space-time,  later for a cosmological background.
For the present purpose, it does not matter that the  degrees of freedom of the models are not ``fundamental'' in the sense of relational quantum gravity\cite{0264-9381-28-15-153002}:  here,  they simply serve to compute  correlations among measurements, in the same way as quantum models used to interpret many  laboratory systems 
({\it e.g.}, refs. \cite{RevModPhys.71.S288,Handsteiner:2016ulx}).

\subsection{Quantum-spin-algebra model}


The following model is developed to provide a concrete worked example of new quantum-geometrical correlations on scales much larger than the Planck length.
The  goal is to develop a quantum model for emergent proper time relationships with a world line in flat space-time.
More specifically, we need a quantum model with operators that describe the  relationships between  time  on different world lines in different space-time directions.
Taking our cue from Wheeler, the quantum states of the  model should live on light cones, meaning that causal relationships in all directions define an exact symmetry.   

Classical proper time  is a scalar, but
causal relationships defined by a light cone are multidirectional. 
In the quantum system, these requirements can be reconciled if states in different spatial directions are entangled.
All observables  in scalar proper classical time should emerge by contracting nonlocal, orientated states in three spatial directions
into a scalar clock operator.

These properties motivate us to model quantum space-time relationships using a  spin algebra,
 instead of the quantized harmonic oscillation of  scalar amplitude usually used for inflation.
The spin model allows a quantitative estimate of new spooky quantum geometrical relationships that cannot occur in standard theory: nonlocal  entanglement among multidirectional temporal states, with the correct (Planck-scale, holographic) number of degrees of freedom, for a region of  any size.

\begin{figure}
\begin{centering}
\includegraphics[width=\linewidth]{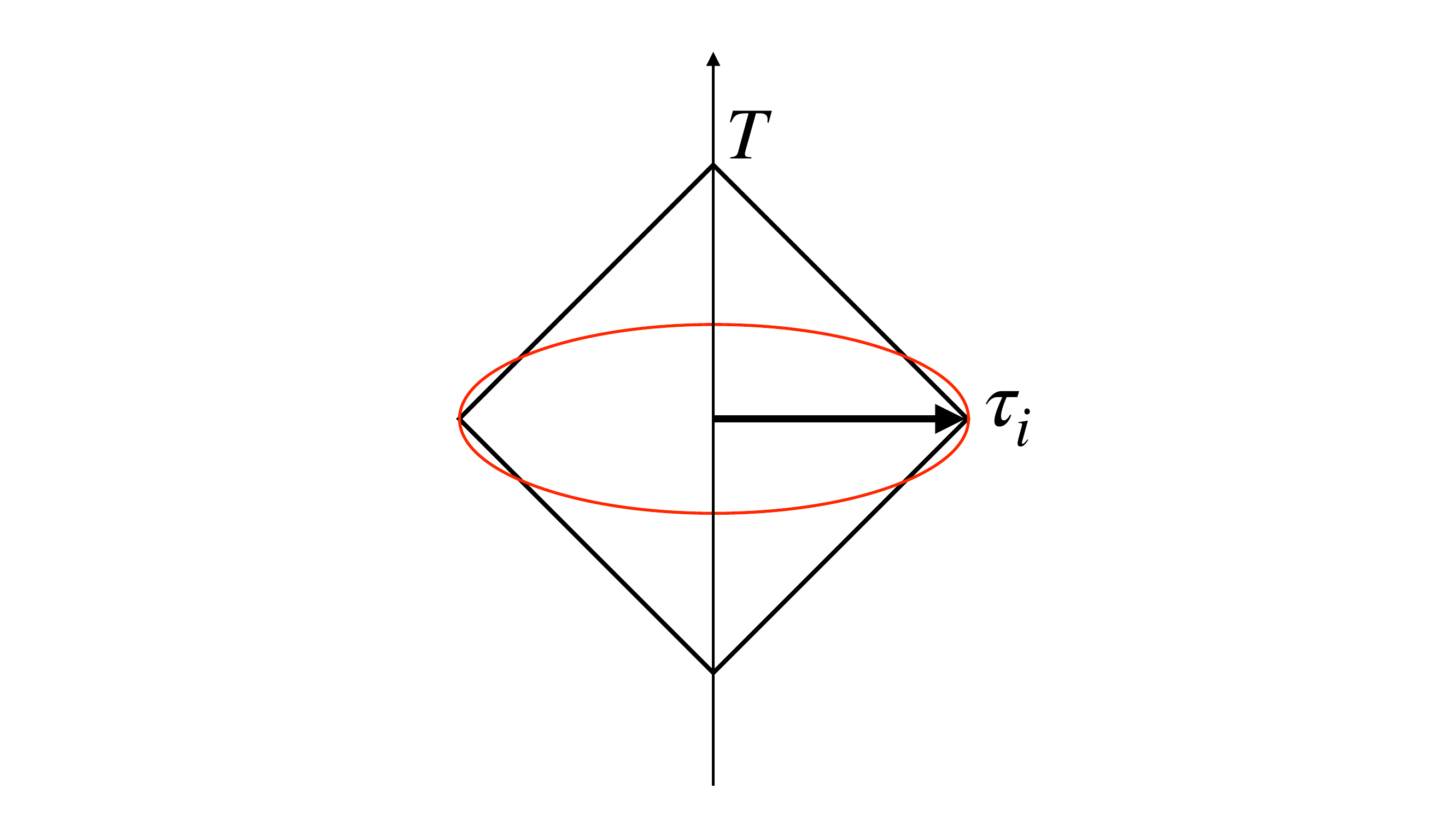}
\par\end{centering}
\protect\caption{Space-time diagram of a causal diamond  associated with an interval on a world line, shown here in the rest frame.  Operators of  spin model quantum states  correspond  to  three noncommuting  directional components of time, $\hat \tau_i$, which combine to form a commuting  operator  $\hat T$,  the total duration along the interval in classical proper time.   They describe positional relationships between an observer and events on the 2D spacelike boundary in different directions.\label{diamonddef}}
\end{figure}

The standard quantum spin algebra is  repurposed here as a  relational holographic quantum model of  a  causal diamond, the region defined by future and past light cones from  an interval of  time on any world line  (Fig. \ref{diamonddef}).  
Fluctuations of the quantum system are  interpreted as geometrical fluctuations of proper time on a world line relative to the 2D spacelike boundary where the light cones intersect--- spooky correlations among directions.
The  temporal correlations on causal diamonds  are  extrapolated below to scalar curvature on inflationary horizons, and ultimately  to distinctive new exotic properties of the matching relic classical perturbations.

The model is defined by quantum operators $\hat\tau_i$ with the dimension of time.
The indices $i,j,k$  take the values $1,2,3$,  identified physically with classical directions in 3-space. 
 The commutation of the operators obey a standard  spin algebra in three dimensions:
\begin{equation}\label{3Dcommute}
[{\hat \tau}_i,{\hat \tau}_j]= it_P {\hat \tau}_k \epsilon_{ijk} ,
\end{equation}
where $ \epsilon_{ijk}$ denotes the Levi-Civita antisymmetric 3-tensor. 
The operators are well known to obey the Jacobi identities
\begin{equation}
[{\hat \tau}_i, [{\hat \tau}_j,{\hat \tau}_k]] + [{\hat \tau}_k, [{\hat \tau}_i,{\hat \tau}_j]] + [{\hat \tau}_j, [{\hat \tau}_k,{\hat \tau}_i]] =0,
\end{equation}
so that  the  quantum theory is self-consistent. 

The quantum operator notation $\hat \tau_i$ is introduced to highlight our unconventional physical adaptation\cite{Hogan2012b} of this familiar system to describe the quantum  entanglement among nonlocal quantum degrees of freedom that emerge as  space and time.     Instead of angular momentum components, the  conjugate variables are directional components of a quantum operator that approximates  time in a classical limit, but has noncommuting relationships among spatial directions.
 With this physical interpretation, Eq. (\ref{3Dcommute}) describes a holographic entanglement of geometrical degrees of freedom over an entire 4-volume.

 In Eq. (\ref{3Dcommute}),    the  Planck time  $t_P$    takes the place of the usual quantum of action, Planck's constant $\hbar$, that governs standard quantum-dynamical relationships associated with displacement operators in a continuous space-time background. As explained below,  the coefficient $t_P$ is chosen so that the number of degrees of freedom agrees with what is needed to produce   holographic emergent gravity as a statistical behavior\cite{Jacobson1995,Verlinde2011,Padmanabhan:2013nxa,Jacobson:2015hqa}. 
 
 The model posits that  quantum spacetime states for a causal diamond  much larger than the Planck time ($T>>t_P$) have the same discrete relationships as quantum states for any high angular momentum system ($|J|>>\hbar$). 
The amplitude, symmetries, and entanglement of fluctuations in emergent time and direction are derived with only  quantum commutators: they do not depend on dynamical operators or a Hamiltonian.  In standard treatments of angular momentum\cite{landau},  the quantum conditions (Eq. \ref{3Dcommute}) are often derived from a correspondence principle  with classical Poisson brackets; here, they are motivated just from their symmetry and holographic information content.

The  spin algebra  combines operators associated with three spatial directions
into a  rotationally invariant algebra.
In this interpretation it  describes a state  in relation to a chosen spatial location, the origin of coordinates, interpreted  as a  clock or observer at rest.
Like an atomic model, the properties of the quantum system are expressed using classical coordinates. The interpretation is extended below to model the emergence of global directions and cosmic time, and the projection of the quantum fluctuations onto  classical cosmological perturbations that arise during inflation.

\subsubsection{Eigenstates of emergent proper time duration}

Emergent  classical proper time duration is described by an operator $\hat T$, analogous to  total angular momentum:
\begin{equation}
\hat T^2 \equiv {\hat \tau}_i^*{\hat \tau}_i \ .
\end{equation}
 In the same way that total angular momentum commutes with all of its components,  
\begin{equation}
[\hat T^2, {\hat \tau}_i] = 0,
\end{equation}
the  emergent proper time duration $T$, the observable defined by eigenvalues  of $\hat T$,  has no quantum uncertainty. 
Causal structure is an exact symmetry by construction:  the radius of the 2D boundary (of the causal diamond) in the observer rest frame is identified   with $ cT$.  Thus, the spin algebra in 3D space actually describes a quantum model of all states in a  4D causal diamond, including the embedded causal diamonds that can nest within it.

Adapting  conventional notation for angular momentum, let quantum numbers $l$ denote positive integers  that label  discrete temporal eigenstates: 
\begin{equation}\label{radialeigenvalues}
\hat T^2|l\rangle =l(l+1) t_P^2|l\rangle,
\end{equation}
corresponding to discrete eigenvalues of classical proper  time duration,
\begin{equation}\label{separation}
T= \sqrt{l (l+1)}t_P.
\end{equation}

\subsubsection{Uncertainty relation for orthogonal directions}

The directional operators $\hat \tau_i$ are related by an uncertainty relation: a  variance $\langle {\tau}_\perp^2 \rangle = Tt_P$ in orthogonal directions that increases with size,  
in  the same way that a  state of definite angular momentum in one direction is a superposition of states in the orthogonal directions.

To show this, consider projections of the operator ${\hat \tau}_i$.
Let  $l_i$ denote its eigenvalues in direction $i$:
\begin{equation}\label{lproject}
{\hat \tau}_i | l, l_i\rangle= l_i t_P |l, l_i\rangle.
\end{equation}
In a state $|l\rangle$, the  operator ${\hat \tau}_i$ can take discrete eigenvalues  in  units of $t_P$,
\begin{equation}\label{positioneigenvalues}
l_i= l, l-1, \dots , -l,
\end{equation}
giving $2l+1$ possible values. 

Still following  standard practice ({\it i.e.}, ref. \cite{landau}), define raising and lowering operators for components in each direction:
\begin{equation}\label{raiselower}
{\hat \delta}_{1\pm}\equiv {\hat \tau}_2\pm i {\hat \tau}_3,
\end{equation}
with equivalent  expressions for  cyclic permutations of the indices.
The effect on a state is to  raise or lower the quantum number of the projection onto that component by one unit (that is, one Planck time),  while leaving the total $T$ invariant. 
In our interpretation, these operators are identified below as discrete, differential,  directional projections on individual light cones  (e.g., Fig. \ref{foliation}), and in the Appendix, as  operators that relate proper time between different world lines.

The duration operator $\hat T^2$ can be written in terms of any single $i$ as    
\begin{equation}
\hat T^2= {\hat \delta}_{i+}{\hat \delta}_{i-}+ {\hat \tau}_i^2+{\hat \tau}_i= {\hat \delta}_{i-}{\hat \delta}_{i+}- {\hat \tau}_i^2+{\hat \tau}_i.
\end{equation}
Direct calculation (e.g., ref.\cite{landau}) then leads to  the following product of amplitudes for  measurements of  either of the   orthogonal components  ${\hat \tau}_j$, with $j \neq i$:
\begin{equation}\label{transverse}
\langle l_i | {\hat \tau}_j | l_i-1\rangle  \langle l_i-1 | {\hat \tau}_j | l_i \rangle = (l+ l_i) (l-l_i+1)t_P^2/2,
\end{equation}
again for any $i$.

\begin{figure}
\begin{centering}
\includegraphics[width=\linewidth]{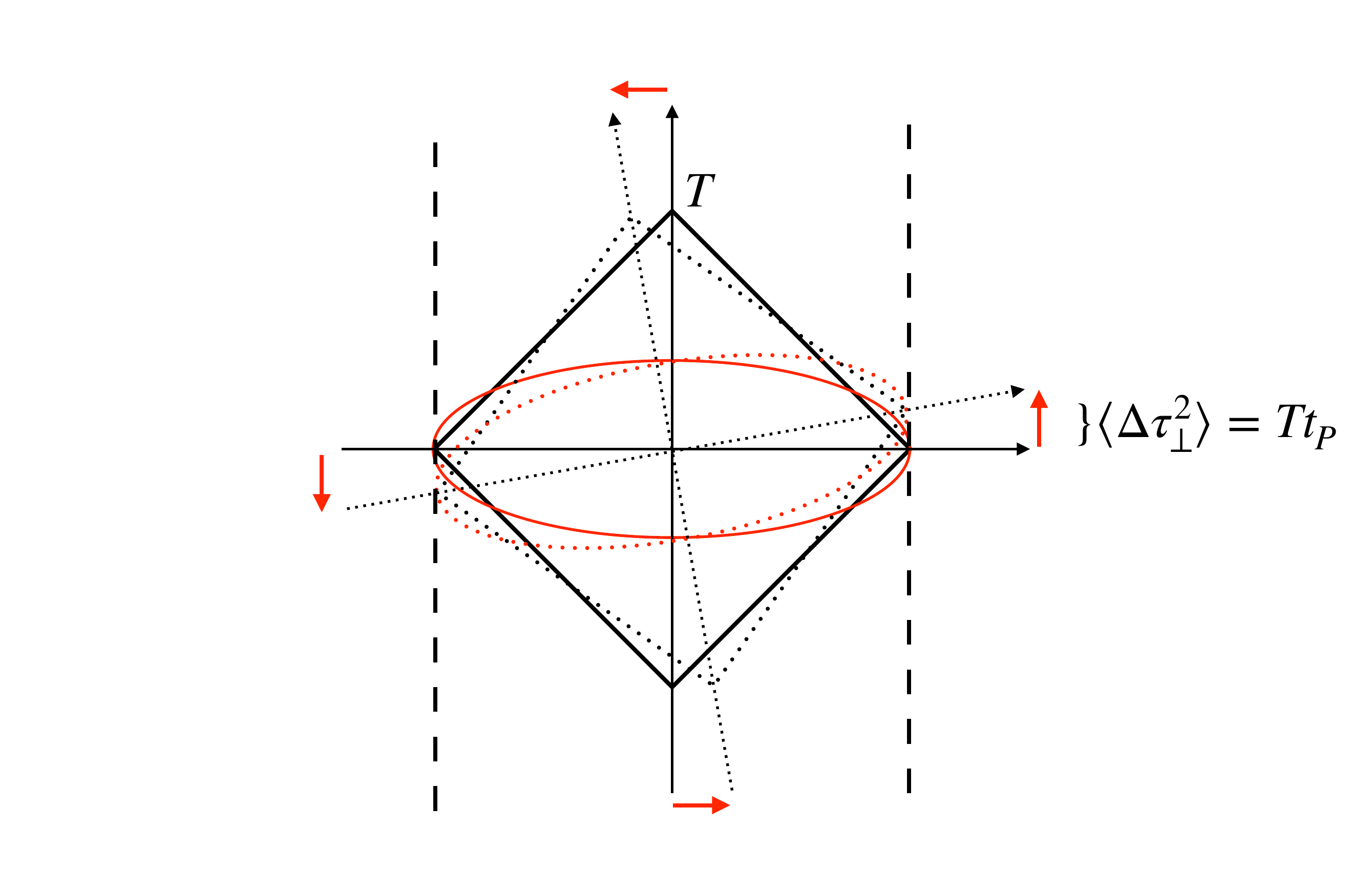}
\par\end{centering}
\protect\caption{Visualization of the quantum spin model fluctuations as  generalized rotations of a causal diamond.
A measurement of time along one axis on the surface of a  causal diamond of duration $T$ and radius $cT$ along any direction (here, out of the page) leads to antisymmetric fluctuations of time on the boundary in the orthogonal directions, of magnitude $\langle \Delta {\tau}_\perp^2 \rangle   =  T t_P$ (Eq. \ref{perpvariance}).
In the spooky model,  antisymmetric fluctuations of  curvature (Eq. \ref{potentialvariance})
 or proper time  (Eq. \ref{anti}) ultimately freeze in as antisymmetric cosmological perturbations (Eq. \ref{antibispectrum}).  \label{diamond}}
\end{figure}

Notice that the left side represents the expectation in an eigenstate $| l_i \rangle$ of an orthogonal-variance operator,
\begin{equation}
 {\hat \tau}_j | l_i-1\rangle  \langle l_i-1 | {\hat \tau}_j .
\end{equation}
Thus, Eq. (\ref{transverse}) gives   the expected variance $\langle \Delta{\tau}_\perp^2 \rangle$ in orthogonal components ${\hat \tau}_j$ in an  eigenstate $|l_i\rangle$ of definite ${\hat \tau}_i$.
This leads to a directional uncertainty relation:   from Eqs. (\ref{separation}) and (\ref{transverse}) in the limit of $l\approx l_i>>1$,  orthogonal  temporal displacements   have  a  variance  about  a mean value $T$ given by
\begin{equation}\label{perpvariance}
\langle \Delta{\tau}_\perp^2 \rangle =  \langle ({\tau}_\perp- T)^2 \rangle  =  T t_P.
\end{equation}
This  relation refers to  time operators in any pair of orthogonal directions, relative to the 2D causal-diamond boundary of radius $cT$  (Fig. \ref{diamond}). 

Thus,  time on the boundary, defined in relation to an  observer at the origin,  is in a superposition of directionally antisymmetric states. 
A causal diamond or horizon surface is never exactly isotropic, but has directionally correlated, antisymmetric fluctuations.

\subsubsection{Physical fluctuations in gravitational potential}

To help clarify the physical interpretation of this strange result, define   operators for displacement
\begin{equation}\label{deltahattau}
\Delta {\hat \tau}_i\equiv {\hat \tau}_i - {\hat T}
\end{equation}
and for  dimensionless fractional displacement,  
\begin{equation}\label{Delta}
\hat \Delta_i  
\equiv \Delta {\hat \tau}_i / T.
\end{equation}
The latter operator represents a difference in potential associated with direction $i$ at separation $cT$.  Fractional time distortions appear as  differences in $\Delta_i$ along the three spatial directions that are all correlated with each other.

As discussed in the Appendix,   the identification of $cT$ with $R$ means that virtual fluctuations in flat space-time are ``paid back'' on the return light cones, for causal diamonds of any size.  Thus,  potential fluctuations associated with measurements on a single  world line  exactly cancel and are not observable.  
The hypothesis of this paper is that during inflation, nonlocal relational correlations resembling those of $\hat \Delta_i$  on different world lines correspond to differences in comoving proper time, or perturbations in scalar curvature in the emergent classical metric on the horizon. This hypothesis has physical consequences.

One physical consequence is a change in the overall amplitude of perturbations.
 During slow roll inflation, the relevant causal diamond radius is  approximately the radius of the horizon, so
the fluctuation power  of dimensionless relic invariant curvature perturbations is approximately given by
\begin{equation}\label{potentialvariance}
\langle \Delta^2 \rangle = \langle {\Delta  \tau}_\perp^2 \rangle /T^2   =   t_P/T = Ht_p.
\end{equation}   
The linear dependence on $H$ is dramatically different from  standard non-holographic field-like perturbations, which scale like $H^2$.
The  basic  reason the geometrical fluctuations are larger than usual is that there are fewer independent degrees of freedom, a direct consequence of  holography.

Another physical consequence is a new directional antisymmetry. In the spin-algebra model, it arises  because $ {\hat \tau_i}$ and  $\hat\Delta_i$ are  odd
 under parity transformations. 
 The statistical properties of global directional antisymmetry in spooky relic perturbations are  derived below  from  invariance  on the classical side.

\subsubsection{Number of eigenstates}

The  eigenvalues of the time operator $\hat T$ are identified  with both classical emergent proper time,  and with the radius of a causal diamond or  horizon of a space-time volume around an observer's world line.
This identification reduces the number of independent dimensions by one.  

The number of  eigenstates within a causal diamond of radius $cT$ can be counted  precisely,   as if they were  discrete angular momentum eigenstates. For each $l$ there are $2l+1$ directional projection eigenstates
so the  number of degrees of freedom ${\cal N}$--- interpreted here as the amount of information or entropy in a causal diamond or horizon---  scales holographically, as the surface area in Planck units:  
\begin{equation}\label{eigenstates}
{\cal N}= \sum_{l'=0}^{l} (2l'+1)\approx  2(T/t_P)^2,
\end{equation}
where the approximation applies in the large $l$ limit, and we have used Eq. (\ref{separation}).
 The  total holographic information  of a causal diamond (Eq. \ref{eigenstates}) counts  all the combinations of nested, entangled light cone states that can represent the state of an interval on the  world line.  
 Up to  factors of order unity in the absolute normalization, this agrees with  the entropy  of black hole horizons.

\subsubsection{Semiclassical visualization as light cone fluctuations}

\begin{figure}
\begin{centering}
\includegraphics[width=\linewidth]{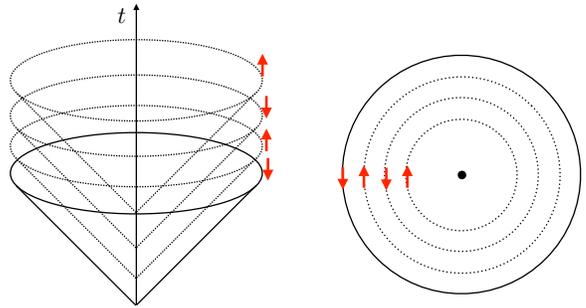}
\par\end{centering}
\protect\caption{Foliation  of flat space-time, adapted from ref. \cite{0264-9381-35-20-204001}. Left side:  a series of light cones separated by a Planck proper time on an observer's world line.  Arrows  indicate  projections of a  raising or lowering operator $\hat\delta_{i\pm}$ along some axis (Eq. \ref{raiselower}).  Right side:  light cones at one  time in the observer's rest frame.\label{foliation}}
\end{figure}

 In a semiclassical picture where  causal diamonds are stitched together from discrete light cones, spooky fluctuations  correspond to Planck-scale differential  displacements on Planck-proper-time-separated light cones  
  (Fig. \ref{foliation}, and refs. \cite{Hogan:2015b,Hogan:2016,0264-9381-35-20-204001}).  Projections of states are directionally antisymmetric  on each light cone, like the rotational raising and lower operators $\hat\delta_{i\pm}$, as discussed in the Appendix.

In an inflationary background (Figs. \ref{horizontwist} and \ref{worldlines}), each light cone imprints a horizon-scale coherent fluctuation when  it  ``freezes'' into a classical metric on the horizon.  
Over an $e$-folding time,  about $ (Ht_P)^{-1}$  null surfaces pass through the horizon.
Each one has a  displacement $\approx t_P$, so the accumulated displacement over a time $1/H$ has a variance $\langle \delta t^2\rangle \approx t_P/H$.  
The curvature  perturbation is   
the fractional time dilation associated with the  fluctuation on the horizon scale over that time, 
\begin{equation}\label{frozen}
\Delta_S^2= \langle \delta t^2\rangle H^2   = \alpha  H t_P,
\end{equation}
where $\alpha$ is a factor of order unity.

This semiclassical fluctuation picture does not fully capture the weird  antisymmetry,  nonlocality and entanglement associated with   the  operators $\hat\delta_{i\pm}$ and $\hat \tau_i$ in the spin-algebra model of relational emergent time.   However, it does lead to the same estimate as 
 Eq. (\ref{potentialvariance}) for  the amplitude: it
depends linearly on the value of $H$ at the time when a fluctuation  freezes  on the horizon.


\begin{figure}
\begin{centering}
\includegraphics[width=\linewidth]{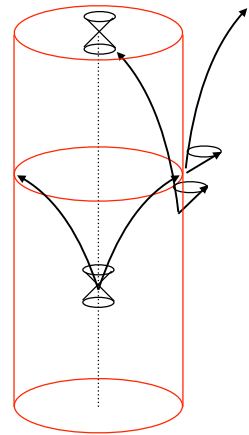}
\par\end{centering}
\protect\caption{
Radial light trajectories in inflation, shown as proper separation from the world line of an observer (dotted line).
During slow roll inflation, the cylinder representing the apparent inflationary horizon, defined by the outermost, incoming null trajectories, lies at approximately constant proper separation from an observer's world line.
It  represents the new quantum-classical boundary shown in Fig. (\ref{penrose}); outgoing states of perturbations freeze in with  nonlocal correlations  on this cylinder.  Inbound light cones of causal diamonds with a boundary near the horizon entangle with outgoing light cones that emerge  significantly later. 
\label{horizontwist}}
\end{figure}

\begin{figure}
\begin{centering}
\includegraphics[width=\linewidth]{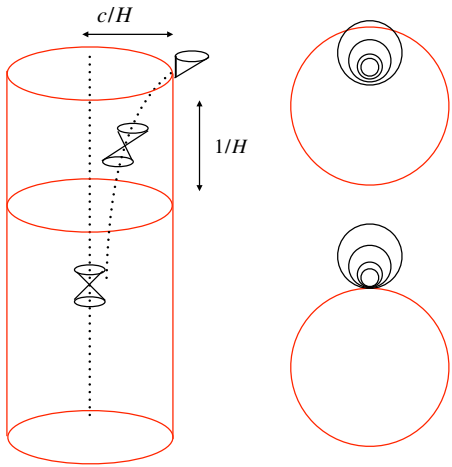}
\par\end{centering}
\protect\caption{In the same coordinates as Fig. (\ref{horizontwist}),  proper spatial separation of two
 comoving world lines (dotted)  are shown with their light cones.
Right side:  Multiple time slices are shown of future light cones of two events, one inside and one on the horizon of the observer.
In the observer frame, clocks appear to freeze  on the horizon.
\label{worldlines}}
\end{figure}

\section{Comparison with current measurements}

\subsection{Constraints from the perturbation  spectrum}

Constraints on the parameters of the  inflationary background model follow  from the result  that the curvature perturbation   $\Delta_S^2$  on any  scale  depends only, and linearly, on the  value of $H$  when it crosses the horizon (Eqs. \ref{potentialvariance} and \ref{frozen}).  Let  $\phi_0$ denote the value of $\phi$ when the measured comoving scales,  comparable to the current Hubble length,  cross the horizon.  From  Eqs.  (\ref{hubble}) and (\ref{frozen}), 
\begin{equation}\label{scalar}
  \Delta_S^2 =     \alpha   ( 8\pi G t_P^2/3)^{1/2} \ \  V(\phi_0)^{1/2}
\end{equation}
The measured value\cite{Ade:2015lrj,Aghanim:2018eyx}  $\Delta_S^2= A_S=  2\times 10^{-9}$ implies an
 energy density during inflation, and an  upper limit to reheating temperature,   characterized by an energy scale  $E_0= V(\phi_0)^{1/4}$  in Planck units:
 \begin{equation}\label{radiationscale}
E_0 \ = \alpha^{-1/2} (3/8\pi)^{1/4}  \Delta_S \ m_Pc^2 \approx  3\times 10^{14}     {\rm GeV},
\end{equation}
where $m_P\equiv \sqrt{\hbar c/G}$.
As usual the actual reheating temperature is generally much less, depending on details of the matter sector.

As in standard inflation, the value of  $H$ is not constant during inflation, but  varies slowly, according to Eqs. (\ref{hubble}) and (\ref{roll}).  Each comoving wavenumber $k$ passes through the horizon at a different time, so the  scalar perturbations vary with scale,  with a  spectrum described by a  spectral index $n_S$:  $\Delta_S^2 \propto k^{n_S-1} $. In the spooky scenario, this ``tilt''  in the spectrum is given  simply by  
\begin{equation}\label{tilt}
n_S-1 \equiv  \frac{d \ln \Delta_S^2}{d \ln k} = \frac{d \ln H}{d \ln k} = - \epsilon,
\end{equation}
where $\epsilon$ denotes the standard  slow roll parameter,
\begin{equation}\label{slowrollepsilon}
 \epsilon \equiv (V'/V)^2 (16\pi G)^{-1} .
\end{equation}
Because $\Delta_S^2\propto H$ and not $H^2$ (as is usual), the tilt differs by a factor of two  from  the standard relation\cite{Akrami:2018odb}.  Thus,  constraints on the allowed potential shape also change: 
potentials preferred in the spooky scenario are strongly excluded 
for standard models, and {\it vice versa}.

Eqs.  (\ref{tilt}) and (\ref{slowrollepsilon}) imply that the measured tilt  depends only on  $V'/V$ at the epoch when the measured  range of scales passes through the horizon. 
The measured value \cite{Aghanim:2018eyx,Akrami:2018vks}  $1-n_S = 0.035\pm0.004$ constrains   its logarithmic slope  to be close to the inverse  Planck mass: 
\begin{equation}\label{slope}
 \left(\frac{V'}{V}\right)_{\phi_0} = \frac{\sqrt{16\pi \epsilon}}{ m_P} = 1.32 m_P^{-1}  \left(\frac{1-n_S}{0.035}\right)^{1/2} .     
\end{equation}
 
As usual, sufficient inflation to reach the current scale of the universe requires $N\approx 60$ e-foldings since $\phi=\phi_0$, depending on reheating and subsequent evolution. 
In the slow roll approximation,   
\begin{equation}\label{Nfold}
|\dot \phi/\phi|_{\phi_0} \approx H(\phi_0)/N.
\end{equation}
Combination of  Eqs. (\ref{hubble}),  (\ref{slowroll}),  (\ref{slope})  and (\ref{Nfold}) leads to an absolute estimate  of 
$\phi_0$:
\begin{equation}\label{phi0}
\phi_0 \approx  \frac{N}{8\pi} \left(\frac{V'}{V}\right)_{\phi_0} m^2_P\approx  3.1 \ m_P \  \frac{N}{60}  \left(\frac{1-n_S}{0.035}\right)^{1/2}.
\end{equation}
These  results show that properties of the effective potential $V$ in the exotic scenario are in principle overdetermined by measurements. The value and slope of the potential  determine respectively the  amplitude and spectral tilt of the relic perturbations.  Given the tilt, the  value of $\phi_0$ determines the number of e-foldings--- that is, the size of the currently  observable universe. 
 
 

It   is not trivial for a potential to satisfy these experimental constraints on both $N$ and $n_S$. For example, a potential of  monomial form $V\propto \phi^b$ satisfies Eqs.  (\ref{slope}) and (\ref{phi0}) if and only if 
\begin{equation}\label{mono}
b = \phi_0 (V'/V)_{\phi_0} = 2N\epsilon  = 4.1 \  \frac{N}{60} \left(\frac{1-n_S}{0.035}\right),
\end{equation}
so cosmological measurements agree (to within measurement errors) with $b =4$, but not with  other integer values.  
A potential of the form
\begin{equation}\label{phifour}
V= {\cal V} \phi^4,
\end{equation}
fits current measurements with $N= 59\pm 7$, and a coefficient
 ${\cal V} \approx 10^{-20}$ in Planck units that  depends on $\alpha$,  $N$ and $n_S$. 
 A  potential of  this  form  (Eq. \ref{phifour})   is now ruled out for standard inflation\cite{Akrami:2018odb}.
 The range of viable models will be more constrained by improved measurements of the tilt.

\subsection{Comparison with inflaton field fluctuations}

In standard slow-roll inflation,  scalar fluctuations $\Delta_{S,\delta\phi}$ from quantum fluctuations of the inflaton field (e.g. \cite{Kadota:2005hv,Baumann:2009ds,Kamionkowski:2015yta}) depend not only on $H$, but also on $\epsilon$:
\begin{equation}\label{standard}
\Delta_{S,\delta\phi}^2 = (H^2/2\pi \dot\phi)^2= \frac{1}{8\pi^2}  H^2 t_P^2 \epsilon^{-1}.
\end{equation}
These effective modes presumably still exist in the spooky system, but they are subdominant.
 Comparing Eq. (\ref{standard}) to Eq. (\ref{frozen}), the exotic effect dominates for observationally viable values of $\epsilon$ and $H$ in the spooky scenario:
\begin{equation}
\Delta_{S}^2/\Delta_{S,\delta\phi}^2= 8\pi^2 \alpha\epsilon /H(\phi_0)t_P >> 1,
\end{equation}
which validates the consistency of our approximation to neglect the gravitational effect of virtual inflaton fluctuations.  The value of $H$ is now so small that  they are unimportant, except for very small values of $\epsilon$.

Similarly, the  exotic scalar spectral index tilt (Eq. \ref{tilt})--- like the standard prediction for tensor tilt---  depends only on $\epsilon$, whereas the standard prediction, because it also depends on the second derivative of $V$,  allows a larger variety of potentials that fit measurements.
The running (change with scale) of the spectral index is predicted to be very small,  as in many standard models.

\subsection{Tensor perturbations}

To agree with current Planck-sensitivity laboratory constraints\cite{holoshear}, an exact symmetry is built into the models here  (for example, in Eq. \ref{3Dcommute}) such that spooky tensor-like correlation modes vanish.  Because of the directional antisymmetry, perturbation multipoles are only odd; the even directional multipoles, including quadrupolar gravitational waves, vanish.   As a result,   tensor perturbations from spooky fluctuations are predicted to be  small.   

Of course, gravitational waves must still exist, but in emergent theories of gravity,  gravitational waves, like curvature, are emergent rather than fundamental degrees of freedom\cite{Jacobson1995,Verlinde2011,Padmanabhan:2013nxa,Jacobson:2015hqa}. The standard theory of a spin-2 graviton has a similar status to the theory of phonons--- they are physically real, but are not fundamental quanta.
Their quantum fluctuations should be  described by the standard  effective theory,  linearized general relativity. 

In the context of inflation, the usual quantum theory of tensor modes applies to these effective degrees of freedom.
The  metric can  be  quantized in the standard way by  linearized quantum  gravity,  so 
tensor perturbations  occur with the standard value, $\Delta_T^2=  H^2 t_P^2/ 2\pi^2$.  The exotic scenario thus predicts  a tensor to scalar ratio
\begin{equation}\label{tsratio}
r =\Delta_T^2/ \Delta_S^2 = Ht_P/ 2\pi^2 \alpha = \Delta_{S}^2  / 2\pi^2  \alpha^2,
\end{equation}
which is  far too small to  measure.  It is many orders of magnitude below the current  experimental upper bound\cite{Array:2015xqh,Akrami:2018vks}, $r<0.07$, and much smaller than predictions of some standard slow-roll inflation  models (e.g. \cite{STAROBINSKY198099,Starobinsky:1983zz}) that fit current data well\cite{Akrami:2018odb} without spooky correlations.

\subsection{Consistency of the Effective Potential}

In standard inflation,   a  ``super-Planckian''  value of $\phi$, as in Eq. (\ref{phi0}),  often leads to inconsistency from divergences in an effective field expansion\cite{Baumann:2009ds,Kamionkowski:2015yta}.  However, in an emergent space-time, this apparent difficulty could be an artifact of inappropriately applied quantum field degrees of freedom: classical space and time are  separable only in systems much larger than the Planck length, and  quantum field degrees of freedom are   separable from space-time only well below the Planck mass.   
In this context, it is consistent to adopt a  classical approximation for  the unperturbed background geometry  on the scale $H^{-1}>>t_P$ with any classical expectation value $\phi $.   

As in many inflation models, the exotic scenario does not 
address the  physical origins of  $V(\phi)$, or its connection with  known matter  fields. The one small number in the model (which can be taken as the coefficient ${\cal V}$ in Eq. (\ref{phifour})) is not explained.

\section{ Signatures of Spooky   Correlations} 

The last section showed that the power spectrum of perturbations in the spooky scenario agrees with standard concordance cosmology, and  with current data.  However,  covariances significantly  depart from standard predictions for some observables:
 unique spooky correlations among relic mode phases
produce measurable statistical signatures in the  distribution of matter and radiation at late times,  that
 distinguish  spooky models  from standard inflationary fluctuations or latter-day classical processes. 

The following  considerations do not rely on specific features of the  quantum models introduced above.
As before,  space-time in the classical era---   above reheating in Fig. (\ref{penrose})--- is described by a standard FRW background metric with linear curvature perturbations. The perturbations are required to obey the usual constraints that apply to any space-time, such as general covariance, as well as the standard global cosmological symmetries of homogeneity and isotropy.   The new feature added by spooky inflation is to relax the usual constraints on locality of  initial conditions.  New kinds of spooky spacelike correlations permit  phase correlations, among classical modes in different directions and on different scales, that are not possible in the standard picture.  
The  new correlations are still highly constrained by cosmological symmetries, and must obey a new directional antisymmetry that is potentially observable.
As elaborated further in the Appendix, in  a fully relational model of quantum gravity this classical relic statistical signature  ultimately corresponds to an antisymmetry of relational quantum states similar to those studied above.

\subsection{Classical  perturbations}

As above, assume  a standard unperturbed classical background cosmology, including  (unquantized)  slow-roll inflation and the standard late-universe concordance model, $\Lambda$CDM.
In linear perturbation theory\cite{PhysRevD.22.1882}, a gauge-invariant  curvature perturbation $\Delta(\vec x)$ is constant with time on a world line at fixed comoving coordinate $\vec x$. 
The  transform  in comoving wavenumber space $\vec k$ is:
\begin{equation}\label{complex}
\tilde{\Delta}(\vec k) =\int d\vec x \Delta(\vec x) e^{i\vec k \cdot \vec x} =  |\tilde{\Delta} (\vec k) |  e^{i\theta(\vec k)}.
\end{equation}
For linear perturbations  with only pressureless cold matter, both the modulus $|\tilde{\Delta}(\vec k)|$ and phase $\theta(\vec k)$ of modes are constant. 
Mean square curvature perturbations  are given by integrals over the power spectrum $\Delta_S^2$.  As discussed above, in the real universe, perturbations are statistically isotropic, and  close to scale invariant:  $\Delta_S^2 \propto |k|^{n_S-1}$,  where $n_S$ is close to 1.

On very large scales today, not only the power spectrum but also the actual distributions  $\Delta(\vec x)$ and $\tilde{\Delta}(\vec k)$ are almost the same now as they were at the end of inflation.  
They are modified by a modest factor by radiation-pressure-driven movement of baryons before recombination,  but even so, until they become nonlinear at late times, the comoving position of the bulk of the matter (that is, cold dark matter) in a large scale mode has moved only a small fraction of a wavelength from where it originated.  We can say that  primordial phases,  still preserved in relic linear perturbations of  density on large scales,  ``remember'' the detailed pattern in comoving coordinates that was impressed by the  process that formed them during inflation.

\subsubsection{Correlations of standard inflationary perturbations}

Cosmic  perturbations in the standard picture arise from the gravitational effect of quantum fluctuations of the inflaton field around its expectation value, frozen in when they cross the horizon during inflation.
In simple models   based on gaussian fluctuations of a free quantum field,  the phases and amplitudes of each mode are independent random variables set by an initial vacuum state. In this case, $\tilde{\Delta}_S^2(|k|)$ contains all the information that remains of the primordial process.

In a broad class of widely studied nongaussian models of locally-interacting fields during inflation, the $|\tilde{\Delta}(\vec k)|$'s can be correlated with each other.
The usual measure of correlations among modes is the bispectrum ({\it e.g.}, ref. \cite{Maldacena:2002vr}),
\begin{equation}\label{bispectrum}
B(\vec k_a)= \langle \tilde{\Delta}(\vec k_1) \tilde{\Delta}(\vec k_2) \tilde{\Delta}(\vec k_3)\rangle,
\end{equation}
defined as an average of  transforms $\tilde{\Delta} (\vec k)$ for  triplets of wave vectors $\vec k_a$ that contribute to the distribution.
It is well known that for correlations from  local field interactions, including nongaussian correlations of fields, the bispectrum  
is nonzero only for a  closed triangle  of wave vectors,  $\sum_a \vec k_a  = 0$.  
This property is associated with local momentum conservation for  interactions.
This  broad class of nongaussian models  has been tested using recent data\cite{PlanckXVII}.  

We will now show that spooky correlations have distinctive properties that are cleanly distinguishable from any of these models.
They  have a gaussian distribution of amplitudes, but  also spooky nonlocal and multidirectional phase correlations that cannot be produced by 
any local field theory on a  classical background.

\subsection{ Spooky Perturbations}

In the spooky model, the emergence from a quantum system of a classical geometry ---  an expanding  universe  with a local cosmic standard of rest---  is inseparable from the formation of perturbations.
Fluctuations in the process of emergence {\it are} the source of the perturbations.

Nonlocal  collapse of the wave function--- the projection of the quantum state onto an emergent observer's comoving frame---  ensures that the states of  nested causal diamonds are consistent. They entangle with each other to approximate a  classical local inertial frame everywhere consistent with the global emergent metric. 

The spooky correlations violate  locality and local  momentum conservation in a particular and highly constrained  way.  
The physical  process must still be generally covariant--- it can only depend on quantities that do not depend  on coordinates  or a particular  classical solution.
Perturbations must  also respect cosmological symmetries on all scales during the classical era after they leave the inflationary horizon--- they must be statistically homogeneous and isotropic. 

On the  other hand, space and time are now slightly indeterminate on  pre-emergent  scales, 
so  local momentum conservation is no longer  imposed by a classical metric (and a local inertial frame) on perturbations. 
Physically, this means that momentum  and emergent  time can be  virtually ``borrowed'' and ``paid back'',  on the scale of the horizon.  It is not necessary for all correlations among non-coplanar modes to vanish, only that appropriately invariant averages do.  
The  statistical properties of emergent spooky perturbations must depend only on covariant combinations of wave vectors. 

\subsubsection{General covariance, statistical isotropy,  antisymmetry}

Consider first the requirement of general covariance.
Let  $u_\phi^\nu$ denote the 4-vector field defined by the timelike  inflaton field gradient, and let $k_1^\kappa, k_2^\lambda, k_3^\mu$ denote a triplet of  perturbation mode wave vectors in 3+1 dimensions.
Using the   antisymmetric  Levi-Civita 4-tensor $\epsilon_{\kappa\lambda\mu\nu} $, define a covariant scalar projection,
\begin{equation}\label{fourD}
{\cal E}_{4D}   \propto  \epsilon_{\kappa\lambda\mu\nu} k_1^\kappa k_2^\lambda k_3^\mu u_\phi^\nu.
\end{equation}
This expression is manifestly invariant under coordinate transformations for any triplet of wave vectors.

Now consider the spatial projection onto standard expanding comoving coordinates in 3D.
As usual, the homogeneous background, encoded here in  $u_\phi^\nu$, breaks boost invariance.
 In the  cosmic comoving coordinate frame,  $u_\phi^\nu\propto (1, 0, 0, 0)$, so  Eq. (\ref{fourD})  projects onto
 surfaces of constant comoving proper time  
as   a  scalar triple product for each  triplet of  3D $\vec k$'s:
\begin{equation}\label{triple}
{\cal E} (\vec k_1,\vec k_2,\vec k_3) \equiv  \epsilon_{ijk} k_1^ik_2^jk_3^k  / k_0^3.
\end{equation}
Geometrically, this  dimensionless  triple product represents the oriented volume of the  parallelepiped defined by the $\vec k$'s.
It  vanishes when the $\vec k_a$'s  lie in the same plane, so any closed triangle maps onto  ${\cal E}=0$.
Up to a choice of normalization scale $k_0$, 
${\cal E}$ represents a unique  invariant number derived from a 3D comoving wave vector triplet. 
It is odd under spatial reflections, $\vec k\rightarrow-\vec k$.

 This projection shows that it is possible  to produce a scalar distribution in comoving space that is generally covariant, statistically isotropic,  and  directionally antisymmetric. For this to happen,  is necessary to abandon the constraint of local momentum conservation that requires coplanar momenta.
 
 The  directional antisymmetry can be traced to how locality arises in an emergent  system, where  positions in relation to an observer arise from  directional quantum operators,  rather than a fixed background. In this kind of system, the odd parity of  3D spatial projections of directional relationships onto comoving 3-space by the inflationary horizon is a generic behavior  that largely follows from classical covariance.   Since a three dimensional space is spanned by three basis vectors, and the scalar product of three non-coplanar vectors is  antisymmetric,
an emergent  scalar with zero mean, such as a spooky curvature perturbation,   is naturally odd under  reflections and vanishes at the origin,
\begin{equation}\label{odd}
\Delta(\vec x)= -\Delta(-\vec x) \  \ {\rm and} \ \  \tilde\Delta(\vec k)= -\tilde\Delta(-\vec k).
\end{equation}
Physically, the antisymmetry arises from a simultaneous antisymmetric ``collapse'' of the wave function at horizon antipodes, as the classical comoving frame forms on the inflationary horizon.
The same  projection properties are familiar in angular momentum, such as those in the quantum spin algebra discussed previously.



 \subsubsection{Measures of scale-invariant antisymmetry}

Antisymmetric projections can be used to define   global statistical  measures of 
 spooky nonlocal multidirectional  correlations that emerge  in  classical perturbations.  For a spooky inflation, the most useful projections are scale-invariant:  the normalization $k_0$ for each triplet has a value such that ${\cal E}$ only depends on the shape of the parallelepiped, not on its absolute scale.

The simplest  covariant, scale-invariant normalization choice for ${\cal E}$ is  $k_0^3=  |\epsilon_{ijk} k_1^ik_2^jk_3^k |$:  
\begin{equation}\label{triplezero}
{\cal E}_0 (\vec k_1,\vec k_2,\vec k_3) \equiv  \epsilon_{ijk} k_1^ik_2^jk_3^k  /|\epsilon_{ijk} k_1^ik_2^jk_3^k |.
\end{equation}
With this choice, 
 ${\cal E}_0=\pm 1$ for all non-coplanar triplets; it just measures their  parity.
 
The parity projection (Eq. \ref{triplezero}) does not 
differentiate correlations among modes of different scales.
Other  scale-invariant normalizations are possible that allow spatial filtering.  They offer several  advantages:  measurements over limited ranges of $|\vec k|$;  
  measurement of correlations among  scales that freeze out at different times;  and
explicit  tests of  scale invariance. 

\begin{figure}
\begin{centering}
\includegraphics[width=\linewidth]{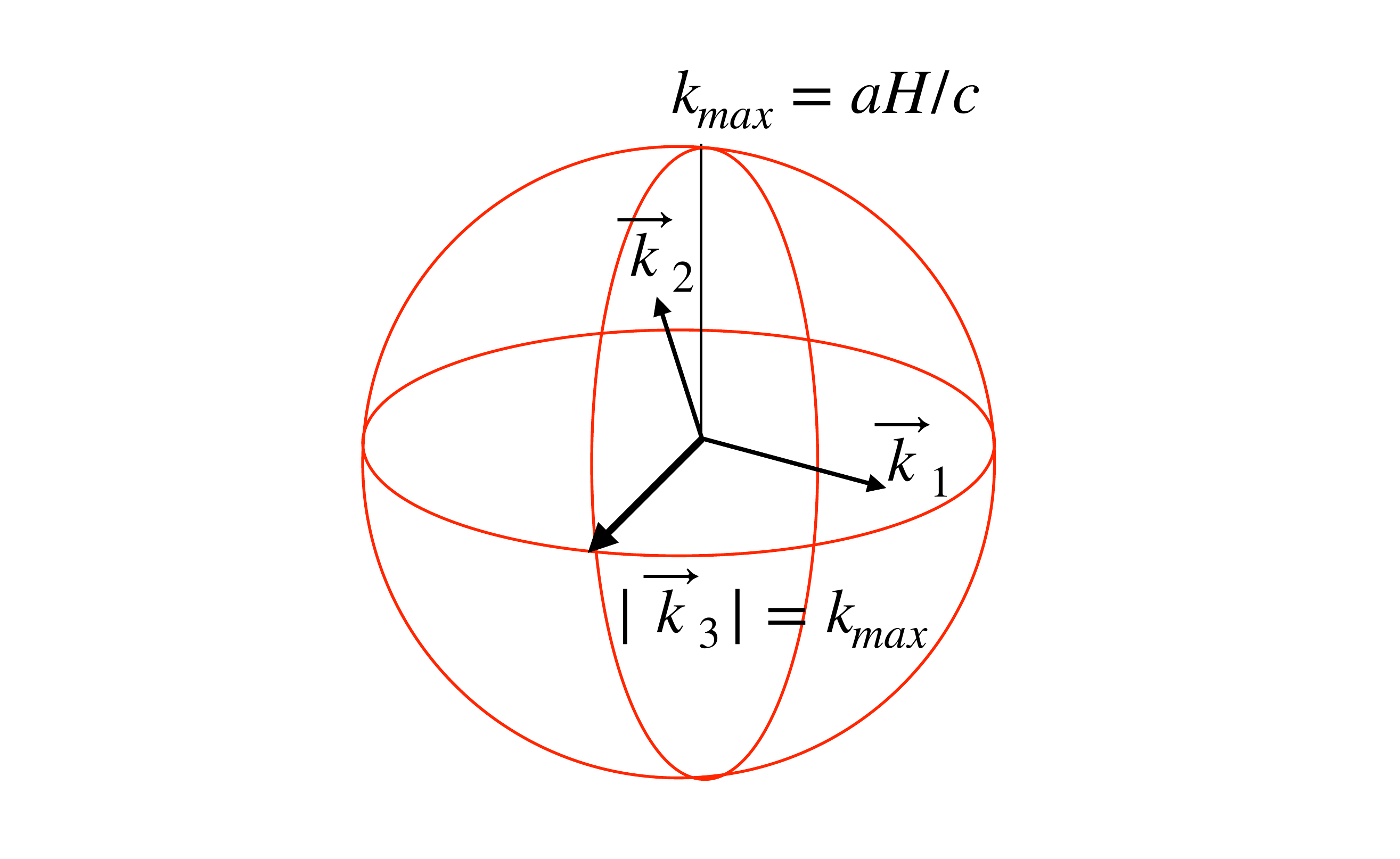}
\par\end{centering}
\protect\caption{ Timing of freezing   of entangled mode states in comoving $\vec k$ space.   A  triplet of comoving wave vectors is shown at a time when the largest of them, the last of the triplet  to freeze out, matches the inflationary horizon, that is,  at $ |\vec k|= aH/c$.       \label{modefreeze}}
\end{figure}

One such projection is motivated by a simplified physical picture, shown in Fig. (\ref{modefreeze}).  
During inflation, the initial conditions for classical mode correlations are determined physically  as entangled states in three spatial directions  freeze on the horizon.
In this  view,  the  state of each mode when it freezes is drawn from a distribution determined by the states of already-frozen modes with smaller $|k|$ in all directions.
The ongoing entanglement of the quantum system ends when the smallest comoving scale freezes in, when the apparent horizon disappears  at the end of inflation--- the end of  the null quantum region boundary in Fig. (\ref{penrose}). 
Over the measured astronomical range of scales, this process is approximately scale-free, and the same should be true of frozen correlations among modes.

A physically natural   scale-invariant normalization for the projection ${\cal E}$ is  thus
the magnitude of the  last mode of each triplet to freeze out, 
\begin{equation}\label{kmax}
{\cal E} (\vec k_1,\vec k_2,\vec k_3) =  \epsilon_{ijk} k_1^ik_2^jk_3^k/ k_{max}^3,
\end{equation}
where
\begin{equation}\label{kmaxdefine}
k_{max}\equiv \max[|\vec k_1|,|\vec k_2|,|\vec k_3|],
\end{equation}
as illustrated in Fig. (\ref{modefreeze}).   In this case,  ${\cal E}$ takes values  -1 or 1 when the $\vec k$'s are orthogonal with equal lengths, and lies between these values if any of the $|k_j|$'s differ.  
This normalization distinguishes the shape   as well as parity of the oriented  parellepiped defined by the vector triplet, so it  allows scale-free statistical measures of the  entanglement between  different scales as well as different directions. 
The discussion below assumes a scale-invariant  definition of ${\cal E}$, of which Eq. (\ref{kmax}) is one example. 
The normalization in Eq. (\ref{kmax}) also allows spatial filtering with just a single scale, which will be used below for practical spookiness estimators.

In the place of Eq. (\ref{bispectrum}), a  new kind of antisymmetric bispectrum can be defined using the invariant antisymmetric projection ${\cal E}(\vec k_1,\vec k_2,\vec k_3)$:
\begin{equation}\label{antibispectrum}
{\cal B}= \langle {\cal E}(\vec k_1,\vec k_2,\vec k_3) \  \tilde{\Delta}(\vec k_1) \tilde{\Delta}(\vec k_2) \tilde{\Delta}(\vec k_3)\rangle.
\end{equation}
It measures correlations of $\Delta$  that are odd under reflections. It is only nonzero for non-coplanar triplets of wave vectors, so it vanishes in standard theories; it is the simplest example of a  measure of ``spookiness'' in the distribution.

\begin{figure}
\begin{centering}
\includegraphics[width=\linewidth]{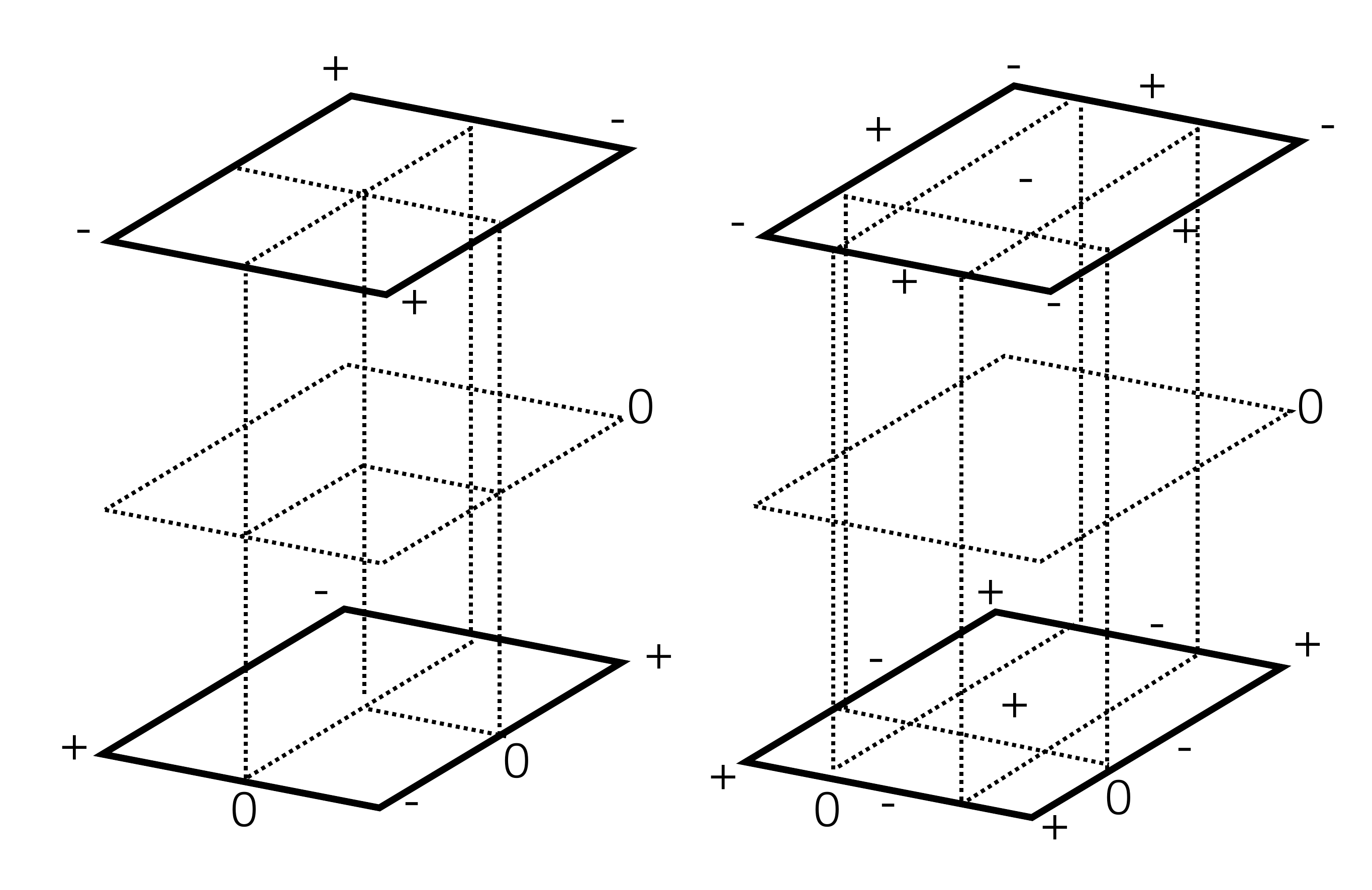}
\par\end{centering}
\protect\caption{ Exploded view in 3-space of cells showing maxima, minima, and zeros of the two types of rectilinear antisymmetric  triplet  eigenmodes  (Eq. \ref{eigen}), for an observer at the center. Some spatial planes of zeros are shown as dotted lines. \label{3Dexplode}}
\end{figure}


%

\subsubsection{Spooky realizations}

 Odd-parity distributions can clearly be realized mathematically by construction. Instead of a general decomposition into independent plane waves (Eq. \ref{complex}),  an odd distribution can be written as a sum of odd-parity  3D triplet modes that depend jointly on  entangled states in three spatial directions:  

\begin{align}\label{eigen}
\nonumber \Delta(\vec x)  = \sum_{\vec k_1,\vec k_2, \vec k_3} &\ 
\alpha(\vec k_1, \vec k_2, \vec k_3) \sin(\vec k_1  \cdot \vec x)  \sin(\vec k_2  \cdot \vec x)  \sin(\vec k_3 \cdot \vec x)
\\   \,+ &\ \beta(\vec k_1, \vec k_2, \vec k_3) \sin(\vec k_1  \cdot \vec x)  \cos(\vec k_2  \cdot \vec x)  \cos(\vec k_3 \cdot \vec x).
\end{align}
The second row  terms  are odd under reflection through the origin but even on reflection in the  plane determined by $\vec k_2$ and  $\vec k_3$. They
single out a direction on each scale,  but the overall distribution is statistically isotropic when averaged over all scales.


%



The spatial layouts of maxima and minima for odd 3D rectilinear  triplet modes are shown in Fig. (\ref{3Dexplode}). 
In  general, an odd 3D triplet mode  can have different values of wave numbers along the three directions.   A spooky odd-parity cosmological distribution is in general composed  of 
a superposition of  such 3D triplet modes, allowing different orientations on different scales. 
The previous analysis shows that the overall distribution can be statistically isotropic, homogeneous and scale invariant, if the distributions of $\alpha$ and $\beta$ are functions only of the combination ${\cal E}(\vec k_1, \vec k_2, \vec k_3)$.


%

A particular triplet decomposition (Eq. $\ref{eigen}$) is not invariant, but is unique to a particular observer, the origin of coordinates (although any given triplet will apply to a discrete set of periodically-spaced observers).
Since the  spatial distribution always vanishes at the origin of coordinates, it should be
interpreted physically as a realization of time distortion or curvature relative to a freely falling geodesic at the origin, not relative to a globally defined,  unperturbed classical background. This assignment  of a relational observable quantity is  in keeping with the emergent character of the whole metric. In our set-up where a horizon is a quantum object, there is no universal, determinate  ``true'' background metric, only one defined in relation to a particular observer and its particular inflationary horizon.
The odd parity is a  remnant of primordial nonlocal correlations that freeze in at spacelike separation from any observer, on its  horizon.

Here then is frozen quantum weirdness on the largest scales: every observer ends inflation with zero total local perturbation. 
Every observer has a different horizon, so the zero point of the potential is observer-dependent; on the other hand,  all observers agree on measurable quantities, such as differences in $\Delta(\vec x)$ between world lines in a realized classical distribution over a region where they can compare measurements.  There is no inconsistency between observers  because there is no way to tell ``who is right''  about an absolute cosmic background frame.  Different observers  look at different but entangled quantum mechanical subsystems that have collapsed into the same classical state where they overlap.



\subsubsection{Generic signature of spooky inflation}
Of course, the  interesting physical question remains whether such weird correlations are actually produced during inflation.
Fortunately, this question may be answered by measurement.
If  spooky  perturbations are the dominant source of primordial perturbations, the  exotic antisymmetric  correlation power  is  a substantial fraction of the total perturbation, so primordial curvature perturbations are expected to show a large effect,
\begin{equation}\label{orderunity}
\langle {\cal B}^2\rangle \approx \langle\Delta^2\rangle^{3}.  
\end{equation}
For  standard field-mode perturbations, the opposite is true: correlations between modes can only exist for coplanar wave vectors, for which ${\cal E}=0$, so the antisymmetric phase power is predicted to identically vanish:
 \begin{equation}\label{zero}
 {\cal B}= 0.
\end{equation}
Thus, a detection of  $ {\cal B}\ne 0 $  can in principle provide a model-independent signature of  spooky primordial phase correlations.

 \subsubsection{Antisymmetric spookiness estimators}

We now turn  to  practical measures of  spookiness  detectable in  cosmic structure.
In the  projection of the  quantum state vector onto the outgoing space, the  choice of observer defines the measurement.
The  triplet mode decomposition in Eq. (\ref{eigen})  displays exact antisymmetry in all modes around only one point, the observer.
On the other hand, there is a statistical tendency everywhere towards odd correlations, which 
becomes conspicuous by convolving the distribution with an antisymmetric kernel. 
This property should  allow the  spookiness of   relic correlations  in the actual universe
to be estimated  from cosmic surveys.

A realistic  estimator of   correlations   can be   designed  with  a response that optimizes the power of a measurement in a particular situation.  
One approach is to use  antisymmetric wavelets ${\cal W}_L$  that probe spooky correlations mainly on 
  scale $L$, with transforms shaped to have relatively  little response to $k_{max}>>1/L$. 
  To search for the spooky effect, and to differentiate it from standard perturbations, it is not necessary for the wavelets to  match  the exact pattern of  primordial correlation; 
it is sufficient that a suitable convolution of ${\cal W}_L$ with $\Delta$ responds to significant fluctuation power with ${\cal B}\ne 0$.

The convolution is most easily written as  a product in transform space.
The  normalized  spookiness  ${\cal S}$ of a distribution $\Delta(\vec x)$ on scale $L$ can be defined via  a normalized wavelet-dependent functional similar to ${\cal B}$,
\begin{equation}\label{measure}
{\cal S}[{\cal W}_L] \equiv \langle \tilde{\cal W}_L({\cal E})\  
 \tilde\Delta(\vec k_1) \tilde\Delta(\vec k_2) \tilde\Delta(\vec k_3)\rangle \  \langle \tilde\Delta_L^2\rangle^{-3/2},
\end{equation}
where   $\tilde{\cal W}_L({\cal E})$ is an odd  function of the  scale-invariant  combination ${\cal E}(\vec k_1,\vec k_2,\vec k_3)$,
with a  filtering scale $L$  imposed  via a UV cutoff  for each triplet at about  $k_{max} \approx 1/L$.  
Again, for standard perturbations, the spookiness vanishes,
because the $ \tilde\Delta(\vec k)$'s are symmetric.  

A simple choice in transform space $\tilde{\cal W}_L$ would be a filtered antisymmetric spectrum linear in ${\cal E}$,
\begin{align}\label{Wcases}
\nonumber \tilde{\cal W}_L  = {\cal E}, &\ 
\, \qquad  k_{max} < 1/L,
\\  = 0, &\ \, \qquad  k_{max} > 1/L,
\end{align}
with some suitable scale-invariant normalization for ${\cal E}$, such as Eq. (\ref{kmax}).
In configuration space, antisymmetric wavelets can be constructed from normalized sums of odd triplet modes as in Eq. (\ref{eigen}), 


\begin{align}
\nonumber {\cal W}_L = \sum_{\vec k_1,\vec k_2, \vec k_3}^{k<k_{max} (L)}  &\ 
\alpha_{\cal W}({\cal E}) \sin(\vec k_1  \cdot \vec x)  \sin(\vec k_2  \cdot \vec x)  \sin(\vec k_3 \cdot \vec x)
\\   \,+ &\ \beta_{\cal W}({\cal E}) \sin(\vec k_1  \cdot \vec x)  \cos(\vec k_2  \cdot \vec x)  \cos(\vec k_3 \cdot \vec x).
\end{align}
For example, a simple two-parameter wavelet  can be designed by choosing $\alpha_{\cal W}\propto  {\cal E} $ and $\beta_{\cal W} \propto {\cal E}$.


%


%
%

The value of  ${\cal S}$ depends on how well the structure of the wavelet matches that of $\tilde\Delta (\vec x)$.  If  
 spooky perturbations dominate the spectrum,  a normalized antisymmetric wavelet ${\cal W}_L$  well matched to the primordial  structure  yields   ${\cal S}$  of order unity.

By not averaging over one of the wave vectors, one can define a measure of the overall asymmetry of the filtered distribution,  associated  with a specific direction and  scale defined by a wave vector $\vec k_D$:
\begin{equation}\label{dipoleprojection}
{\cal D}_L(\vec k_D)= \langle \tilde{\cal W}_L (\vec k_D,\vec k_2,\vec k_3) \tilde{\Delta}(\vec k_D) \tilde{\Delta}(\vec k_2) \tilde{\Delta}(\vec k_3)\rangle_{\vec k_2, \vec k_3}
\end{equation}
On any given scale $|\vec k_D|$, the directional map  ${\cal D}_L(\vec k_D)$ has an angular distribution given by a sum of only odd-parity spherical harmonics, $\ell= 1,3,5\dots$. Its  dipole ($\ell=1$) mode defines a preferred axis and direction, frozen in at the time  when that scale froze out, $ aH/c\approx |\vec k_D|$.

 A particular realization  breaks directional and translational symmetry on each scale.  
On any  smoothing scale $L$, there is a dipole component with a principal  axis on any larger length scale, defined from Eq. \ref{dipoleprojection}, with  an origin-dependent spatial phase. However, the mean dipole of the  distribution vanishes when averaged over all of space because its direction changes sign from place to place, and the system as a whole is statistically isotropic, because the  principal
correlation axis varies  with scale. The  radial and angular  correlations among  odd directional multipoles,  and correlations between principal correlation axes at different places, depend on how freezing on the inflationary horizon correlates structures over a range of time. Principal axes  for nearby world lines (closer than  $\approx 1/|\vec k_D|$) tend to be aligned, and multipole components around each world line are correlated with each other.

\section{ Measurement  in Cosmic Surveys}

\subsection{Cosmic  Microwave Background Anomalies}

The pattern of CMB temperature anisotropy on large angular scales has long been known to display several surprising ``anomalies''\cite{Ade:2015hxq,Schwarz:2015cma} that are often dismissed as insignificant  flukes.  We now suggest that some of them can instead be naturally attributed to   
spooky inflationary correlations, even though the  perturbation power spectrum is the same as standard $\Lambda$CDM  cosmology. 
The large angle anisotropy approximates  a particularly simple  projection that preserves the antisymmetry of primordial perturbations, so general arguments can 
demonstrate  specifically  how new spooky covariances  modify standard predictions.

As shown above (Eq. \ref{odd}),  spooky inflation predicts directional antisymmetry of  $\Delta(\vec x)$  around any observer.
The CMB temperature perturbation $\delta T/T$ smoothed on large angular scales  is  approximately proportional to  the primordial distribution  $\Delta(\vec x)$ on a sphere,  the cosmic last scattering surface.  
Thus to a first approximation, the  spherical-harmonic decomposition  of temperature anisotropy 
at low angular wavenumbers $\ell$  should be mostly directionally antisymmetric.  The fluctuation band power  ($TT$) should have approximately twice the noise power usually expected in odd spherical harmonics $\ell=1,3,\dots$, and a small fraction of the usual power in  even harmonics $\ell= 2, 4\dots$ 
(This odd/even anomaly is not expected to appear in  temperature-polarization ($TE$) correlation, since  polarization is generated physically by an even-parity quadrupole, so it starts  spatially out of phase with $T$ and $\Delta$.) 
 Although the  intrinsic dipole ($\ell =1$) is predicted to be larger than usually expected,  it is not measured, because it cannot be separated from  the much-larger-still local kinematic dipole. Since  the quadrupole ($\ell=2$)  amplitude is predicted to be small compared with its usually expected value,  
 the octopole ($\ell=3$) modes are  the  lowest harmonics predicted to  contribute appreciable observable anisotropy.

This simple prediction appears to be confirmed: a highly suppressed quadrupole and  prominent octopole modes  are well-established  features in the  $TT$ band-power spectrum\cite{Ade:2015hxq,Schwarz:2015cma}.
The same effect manifests as another well-established anomaly, a remarkably small  two-point correlation function of $\delta T/T$  at  large angular separation\cite{Copi:2008hw,Schwarz:2015cma}. In the spooky scenario,  the dipole subtraction is predicted to remove most of the  fluctuation power on angular scales larger than the octopole.

The odd/even  effect  is found to extend well beyond the lowest multipoles:
it has been estimated\cite{Ade:2015hxq,Schwarz:2015cma} that a statistically 
anomalous excess fluctuation power in odd multipoles  extends up to about  $\ell\approx 30$.  
This also agrees with a simple estimate of  the expectation from primordial antisymmetry: 
at about this angular scale, the temperature anisotropy  is significantly modified from its primordial antisymmetric pattern by waves in the  recombination plasma.

The observed CMB  temperature perturbation comes  from  perturbations in both radiation temperature  and gravitational redshift. On   scales much larger than the horizon at recombination,  both effects  preserve their primordial phase, and they tend to cancel each other.  On smaller scales,  propagating acoustic waves of baryon/photon plasma  change the phase of the radiation temperature relative to the dark-matter-dominated potential.  On scales where the effects reinforce each other, they form the well-known baryon acoustic oscillation  peaks in the angular band power spectrum, the first of which peaks at $\ell\approx 200$. 
The symmetric baryon-photon waves should erase  primordial antisymmetry above the scale where the red wing of this first  acoustic peak matches the average large-angle temperature anisotropy band power. This occurs at about $\ell \approx 30$, the scale  
where parity asymmetry is indeed found to  diminish\cite{Ade:2015hxq,Schwarz:2015cma}.



As seen above, the multipole directions in spooky inflation are not independent. 
Spooky directional correlations could also naturally produce octopolar planarity, and   alignments among  normally-uncorrelated multipole components, as observed\cite{Schwarz:2015cma}.

More rigorous comparisons with measurements are possible, by using standard  linear theory for the post-inflation evolution including baryon and radiation transport, and customizing statistical tests to spooky predictions.  
The relatively small number of independent modes on large scales  limits
 the  statistical  power of  large angle $TT$ anisotropy to test  models---the  $p$-values of the anomalies just described are  typically at the percent level, and in the best cases about ten times smaller\cite{Schwarz:2015cma}--- 
 but more powerful tests of this interpretation might be possible as data improve on  large angle polarization\cite{Copi:2013zja,Yoho:2015bla}.
 

\subsection{Spooky correlations  in galaxy surveys}

\subsubsection{Advantages of 3D surveys}

Large-angle CMB anomalies hint that  spooky primordial correlations may indeed have an antisymmetric structure.  If so, the idea can in principle be tested with
 3D surveys of cosmic structure, which  have  more information and   statistical power than CMB surveys. They   contain  many more modes, since they can  measure linear primordial correlations  in  a large 3D volume,
on scales much  smaller than the  horizon.   

In addition,  3D density structure preserves primordial phase information over a wider range of scales than CMB temperature does.
As noted above,  primordial perturbations even in the linear regime are  modified by an early nongravitational effect:  the acoustic propagation  of baryon-photon waves before recombination shifts the phase of  baryon density modes, and symmetrically randomizes  their phases  on scales up to  approximately the horizon scale at recombination, effectively erasing $TT$ antisymmetry.  Since  the  the baryons constitute only about a fifth of the total matter density,  the baryonic oscillations have a relatively small effect on the primordial  spatial distribution of the potential. They are neglected in the rough estimate   given here of  survey sensitivity.

\subsubsection{Spookiness estimators based on linear density contrast}

A  survey of galaxies or gas provides an estimate of  mass density  $\rho(\vec x)$ when convolved over some  kernel with a smoothing scale $L$.  
In the linear regime, the density contrast $\delta \rho(\vec x)\equiv \rho(\vec x)- \langle\rho\rangle $  is proportional to  $\Delta(\vec x)$  with the same kernel,  with a  linear coefficient that depends  on  $L$:
\begin{equation}\label{density}
(\delta \rho(\vec x)/\langle\rho\rangle)_L\approx - \Delta_L(\vec x)  (L_H/L)^2,
\end{equation}
where
 $L_H$  denotes the Hubble scale, $\approx 4000 \rm Mpc$ in the present universe.
 
 In the linear regime, the perturbation in potential is  approximately constant with time on each scale.  
  While the density contrast is not constant, the primordial pattern in comoving space is  approximately preserved,  until  the density perturbation becomes nonlinear.  
 A 3D galaxy survey thus allows   a spookiness estimator similar to Eq. (\ref{measure}),  based on density contrast: 
\begin{equation}\label{measuredensity}
{\cal S}_L  = \langle \tilde{\cal W}_L({\cal E}) \tilde{\delta\rho}(\vec k_1) \tilde{\delta\rho}(\vec k_2) \tilde{\delta\rho}(\vec k_3) \rangle \  \langle\tilde{\delta\rho}_L^{2}\rangle^{-3/2}.
\end{equation}
Because all standard models predict directional symmetry,  a directionally antisymmetric wavelet  $\tilde{\cal W}_L$, filtered at   $k_{max} \approx 1/L$ for any  smoothing scale $L$, provides a model-independent spookiness test.  If the  wavelet is well matched to the structure of  dominant spooky correlations, ${\cal S}_L= {\cal O} (1)$; for standard perturbations, ${\cal S}_L= 0.$

\subsubsection{Estimate of survey requirements}

The next  question is whether imperfect measurements of the  linear density field can in principle show evidence of ${\cal S}_L\ne 0$, and  hence signify ${\cal B}\ne 0$ in  primordial  fluctuations.  
The intrinsic limit of sensitivity for many  finite-volume realizations with the same correlations can be written as an estimation noise error,
\begin{equation}
 \delta {\cal S} \equiv \sqrt{ \langle ({\cal S}_{estimated}- {\cal S}_{true})^2\rangle_{realizations}},
\end{equation}
where ${\cal S}_{true}$ refers to the spookiness of the ``true'' primordial comoving linear density field.
For a wavelet optimally matched to the spooky structure,
$ \delta {\cal S}^{-1}$ gives an estimate of the best possible significance of a detection.

Even with an optimal sampling wavelet, there are
 unavoidable noise sources that contribute to $ \delta {\cal S}$:
 nonlinear physical effects associated with galaxy formation that change the mapping of $\Delta(\vec x)$ to galaxy density on small scales, and  $\sqrt{N}$  noise in measurements of  density from a limited sample. We will use  order-of-magnitude estimates of these noise sources as a rough guide to estimate maximum survey sensitivity.

On small length scales where clustering is nonlinear, movement of matter smears out the one-to-one mapping between  primordial potential perturbation and matter distribution: 
the antisymmetry of the primordial pattern gets mixed away on a gravitational timescale  by nonlinear dynamics of orbital motions.  As a result, most of the cleanly recoverable primordial phase information   comes from a  scale somewhat but not too much larger than the scale  where 
density perturbations become nonlinear--- 
roughly the  scale of visible structures of the  cosmic web, such as  voids, pancakes and filaments.

Let $L_*$ denote the smallest scale  where the primordial pattern of curvature perturbations is mostly intact. 
Density contrast has unit variance in $\approx 20$ Mpc diameter spheres, 
so for rough estimation we adopt a scale about twice as large,  $L_*\approx 40$ Mpc, or $L_*/L_H\approx 10^{-2}$.
 
Nonlinear variations on scale $L_*$
add   noise to  measurements  on larger scales.  White-noise density variance in a volume of  size $L> L_*$  scales  roughly  like $ (L_*/L)^{3}$, so the  spookiness measurement noise on scale $L$, 
\begin{equation}\label{whitenoise}
\delta {\cal S}_L \approx   (L_*/L)^{3/2}  \langle (\delta \rho/\rho)^2\rangle^{-1/2}_L \approx (L/L_*)^{1/2},
\end{equation}
is always greater than unity in a single $L$-size volume. The primordial pattern in this sense is fundamentally buried in noise.

Even so, in principle a coherent spookiness signal can be still be extracted from  a large survey volume $L_S^3$ with about  $(L_S/L)^3$ samples on scale $L$;  
the maximum signal to noise ratio scales  like
\begin{equation}\label{SNR}
 [\delta{\cal S}^{-1}] \approx (L_S/L)^{3/2}  (L_*/L)^{1/2}\approx  (L_*/L)^{2}  (L_S/L_*)^{3/2}.  
\end{equation}
This  estimate accounts only for the purely ``geometrical noise''--- the information limit imposed by nonlinear structure.
It errs on the optimistic side: it is the best one could hope for, if the primordial signal is maximally  conspicuous and minimally contaminated.

One  straightforward conclusion  is that the mean square sensitivity   is at most the number of effective  voxels in  the survey volume:
$ [\delta{\cal S}^{-2}]_{max}< (L_S/L_*)^{3}$.
Thus, the survey should  have the largest volume 
possible, $ L_S \approx L_H$.

The steep dependence  $\propto L^{-2}$  in Eq. (\ref{SNR}) shows that most of the signal comes from the smallest measured  structures where the primordial phase survives--- both because there are more structures (or modes), and because the measured quantity, density contrast,  is larger on small scales.  
For optimal sensitivity,  the  map of  density structure should resolve  the nonlinear clustering scale $L_*$.  Expressed in terms of  redshift $\delta z \equiv \delta L H/c$, ideally the resolution in all three dimensions should be better than
$\delta z\approx L_*/L_H \approx 10^{-2}$.
The maximum possible signal to noise ratio degrades quickly if the resolution is poor:
\begin{equation}\label{rough}
[ \delta{\cal S}^{-1}]_{max}    \approx  1000  (10^2 \delta z)^{-2}.
\end{equation}

In addition to the survey volume and resolution requirements, there must be enough galaxies  so that  the sampling-noise contribution to the measurement error $\delta{\cal S}$ is less than the geometrical noise.  Resolving the phase relationships of perturbations in 3D requires at least an order of magnitude more galaxies than simply measuring the direction-averaged power spectrum, which has been the design goal of most surveys to date.  Guessing that measurement of a dipolar density wavelet fit  in three dimensions requires at least a few galaxies along  each direction in each  $L_*$  volume, or perhaps  $ 10^2$ galaxies, the total number of galaxies $N$ in  a Hubble-volume survey must be more than about
\begin{equation}
N\approx 10^2 (L_H/L_*)^3 \approx 10^8.
\end{equation}
With more galaxies, finer details of the primordial correlations can be resolved.
Galaxy-sampling noise scales with $L$ in the same way as the geometrical noise, so this requirement is approximately independent of $L$. 

The optimal sensitivity estimate (Eq. \ref{rough}) is   promising enough to warrant more  study with simulated realizations of surveys and estimators.  
A comparison between realizations with  random initial phases,   and realizations with spooky initial perturbations, can model and bound effects such as nonlinear clustering, numerical artifacts, survey geometry,  sample selection, and nonuniform radial resolution.

\subsubsection{Implementation in Real Surveys}

The current dataset that comes closest to satisfying the above requirements is the Dark Energy Survey (DES)\cite{Drlica-Wagner:2017tkk}.  It  includes  more than $10^8$ galaxies spread over about a Hubble volume  as required;  however, as it is a broad band photometric survey,  it does not achieve $\delta z_* = 10^{-2}$  in the radial direction for the bulk of its galaxies\cite{Hoyle:2017mee}.  Even allowing for this and additional numerical factors that may reduce overall significance  by more than an order of magnitude below the value in Eq. (\ref{rough}),   it is plausible that  DES might achieve $\delta{\cal S}<<1$--- that is, good enough for  a  detection if $|{\cal S}| = {\cal O}(1)$.
DES  may be the first survey capable of discovering spookiness at high significance.

Detailed studies of spookiness would place demands on surveys  beyond design goals of existing and planned projects;
the expanded scope could motivate extensions and possibly new surveys.
In the future, Large Synoptic Survey Telescope\cite{Ivezic:2008fe} will improve on DES in all respects, but  will still not achieve optimal 3D resolution and sampling.   
An optimal survey  would need  good redshift precision, $\delta z < 10^{-2}$,  in a  Hubble-volume, densely sampled survey, with $N\approx 10^{9}$ galaxies.  The largest volumes may some day be mapped at sufficient resolution  using line emission from gas that is not resolved into galaxies.

 \section{summary}
 
Nonlocal, holographic, entangled states on the inflationary horizon, similar to those invoked to resolve black hole information paradoxes, can produce  correlations in relic perturbations observably different from standard inflation models. Many of their properties are fixed by a single scale, the inflation rate $H$ in Planck units, and well known symmetries of the emergent classical background.

The simplest generic consequence of spooky inflation is a nearly scale free spectrum of curvature perturbations, with an amplitude $
\Delta^2\approx H t_P$ significantly larger than those associated with inflaton field fluctuations.  
Application of standard inflation theory with current measurements then yields  direct  constraints on the value of $H$  and the slope  of the effective potential. The shape of the effective potential is constrained to be close to $V(\phi)\propto \phi^4$ in the range of $k$ observed, with  a definite inflaton value several times the Planck mass.  
These parameters for the potential are  ruled out in standard inflation\cite{Akrami:2018odb}.   
Primordial tensor perturbations are predicted to be very small,   based  both on general symmetry arguments,  and on existing Planck-sensitivity laboratory  constraints.

Another distinctive and robust new prediction, in the sense of being insensitive to the details of specific spooky models,  is an exact directional antisymmetry of  the primordial distribution of curvature perturbations,  traceable directly to the nonlocality and directional anticorrelation of  initial conditions  on the horizon, which is forbidden in standard models. Although the gaussian distribution and predicted evolution of the linear power spectrum are unchanged from the standard  $\Lambda$CDM late time cosmological model, primordial antisymmetry will change covariances for some observables,   modifying estimates of  cosmological parameters and tests of consistency.

Signatures of primordial antisymmetry  already appear to be measured in CMB anisotropy, and if they are indeed  due to nearly-scale-invariant primordial spookiness,  they should also  be observable in large scale 3D galaxy surveys, possibly even in existing data. 
Evidence for spooky correlations could signify a dominant role for new Planck scale quantum degrees of freedom in creating cosmic structure, and lead to  empirical  studies of emergent quantum gravity.

  \begin{acknowledgments}
 This work was supported by the Department
of Energy at Fermilab under Contract No. DE-AC02-07CH11359.  
\end{acknowledgments}

 \bibliographystyle{apsrev4-1.bst}

\bibliography{spooky}

\section{Appendix}

\subsection*{Correlations of emergent proper time  between separate world lines}

The spooky inflation scenario is predicated on the idea that space-time emerges from a quantum system.  Basic conceptual elements of classical space-time relationships, such as localized events and local inertial frames,  are approximate, emergent properties of a quantum system with new, exotic correlations.   Although there is no accepted theory of relational quantum gravity, some  properties of the spooky correlations can be guessed from known causal symmetries of the classical space-time.

One model of quantum departure from classical behavior used to illustrate spooky  correlations is the  spin-algebra model of  Eq. (\ref{3Dcommute}), which  describes a nonlocal spatial antisymmetry of  proper time displacement operators on the surfaces of  causal diamonds.
 A simple extension of the model is sketched here  
to connect it with the  relationship of proper time  between separate world lines, encoded in the entanglement  of their states. The relationship is encoded as pure entanglement information, in the form of  an imaginary cross spectrum of  time displacements.

Consider cross correlations between states of light cones on two world lines, $A$ and $B$, that are classically at rest with respect to each other.
Let $ {\hat \delta }_{AB}^\pm$    and  ${\hat \delta }_{BA}^\pm$ denote  operators analogous to the raising and lowering operators $ {\hat \delta}_{i \pm}$  (Eq. \ref{raiselower}):   single-quantum, Planck-scale projections,  in the  $A$ and $B$ rest frames,
along the  $AB$ and $BA$  spatial separation directions respectively. 
In addition to the spatial antisymmetry already established, they are  also odd on reflection in time,  depending on the orientation towards the past ($-$) or future ($+$): 
\begin{equation}\label{timeparity}
{\hat \delta }_{AB}^+ = -   {\hat \delta }_{AB}^-.
\end{equation}
These operators can be used to make a  model of emergence:   eigenvalues of $ {\hat \delta }_{AB}^\pm$, ${\hat \delta }_{BA}^\pm$ represent  projections, on each world line, of states on a discrete Planck time series of causal diamonds, i.e. time intervals, on the other.   

 \begin{figure}
\begin{centering}
\includegraphics[width=\linewidth]{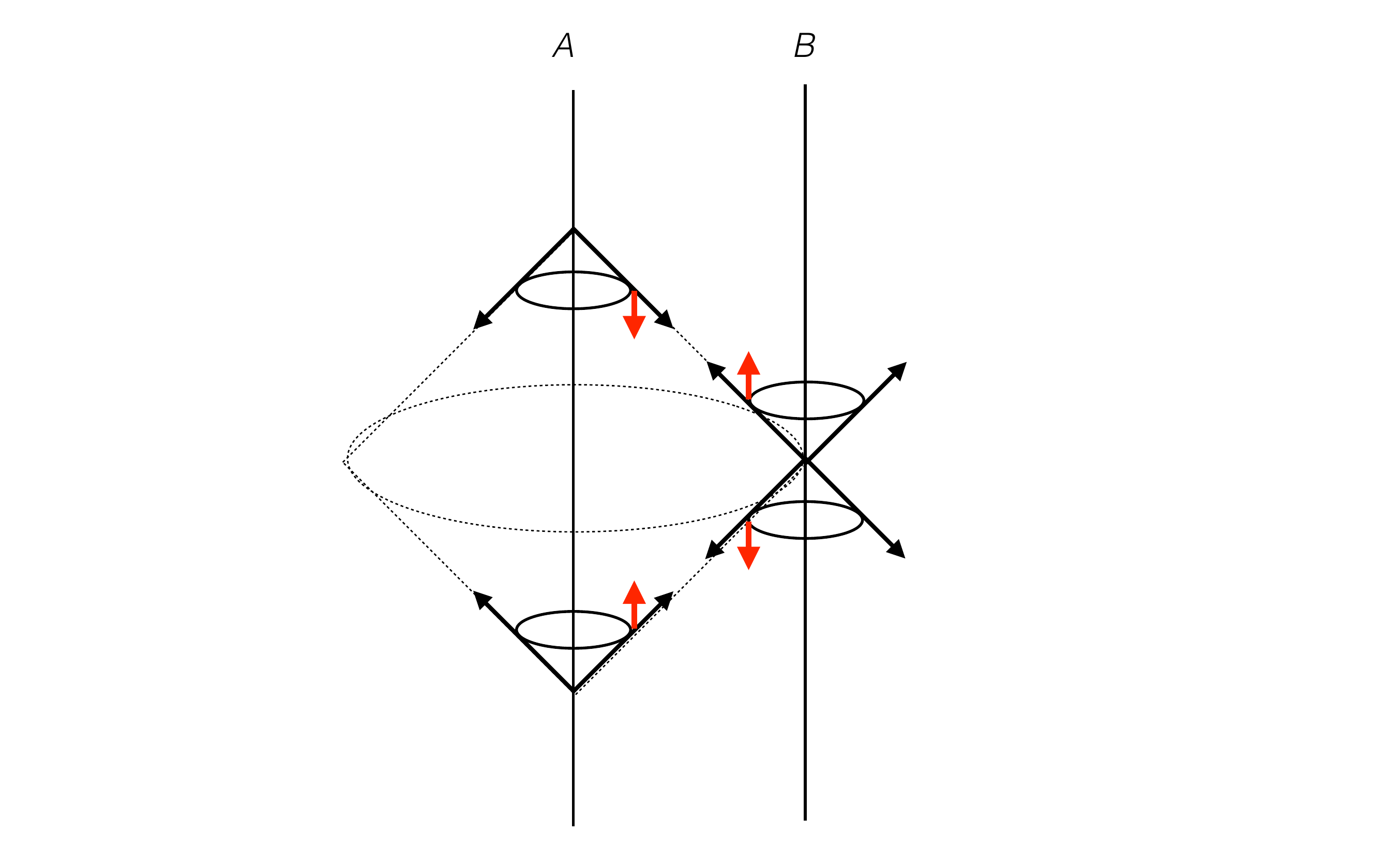}
\par\end{centering}
\protect\caption{Causal relationship of entangled  light cones for two world lines $A$ and $B$ in flat space-time,  adapted from ref.\cite{0264-9381-35-20-204001}. Eigenvalues for these light cones of the relational antisymmetric phase displacement operators (Eq. \ref{timeparity}),  $\hat\delta^{\pm}_{AB}$ and  $\hat\delta^{\pm}_{BA}$,  are shown schematically by arrows. 
The causal diamond state in $A$'s frame describes exotically-cancelling virtual past and future displacements, associated  with the $AB$ direction,   that appear as time-odd displacements in $B$'s frame (Eq. \ref{anti}).    The  cross correlation of causal diamond states describes virtual quantum fluctuations in the relationship of proper  time between the world lines\cite{Hogan:2015b}. 
\label{flatspace}}
\end{figure}

Let  $ \delta_{AB}(t)$ and  $\delta_{BA}(t)$ denote the time series of discrete projections of these operators onto a common classical emergent time variable $t$, again associated with the $AB$, $BA$ directions.   They correspond physically to  combinations of  noncommuting operators that measure in the orthogonal directions, as discussed below.
Each series represents a realization of quantum noise with  a Planck spectral density,  with one bit of information per Planck time. 
Since the states of the two world lines are entangled, the time series are not independent.
As illustrated in Fig. (\ref{flatspace}) for a  causal  diamond on an interval defined by two times on $A$ in flat space,  realizations of the time series  on the two world lines relate to each other with a spooky nonlocal correlation:
\begin{equation}\label{anti}
2\delta_{BA}(t) =  \delta_{AB}(t- R/c) -     \delta_{AB}(t+ R/c),
\end{equation}
where $R$ is the separation. The same relation applies with $A$ and $B$ reversed.

Eqs. (\ref{timeparity}) and (\ref{anti}) express the idea that 
virtual time displacements of $A$ relative to  $B$ represent Planck scale fluctuations of ``borrowed time'' that are ``paid back'' after  a round trip light crossing time.  
It is the counterpart of the  antisymmetry of the nonlocal  operators ${\hat \delta}_{i\pm}$ and $\delta {\hat \tau}_i$,   assigned to  causal diamonds for different observers in the same space-time (Eqs. \ref{raiselower} and \ref{deltahattau}).

This relationship between time series leads to a purely imaginary cross spectrum  in the frequency domain,
\begin{equation}\label{eqn:crossimag2}
\,\tilde{\delta}_{BA} (f)\, 
= \,   i \,  \sin (2\pi f R/c )\ \,    \tilde{\delta}_{AB}(f).
\end{equation}
The cross spectrum between the world lines is imaginary because the cross correlation represents pure entanglement information--- it is not visible in the autocorrelation of either time series with itself, only when the two are compared. The offset phase between them is always 90 degrees, but the actual phase is determined by the state preparation--- in this case, environmental information associated with the states of the  spatial directions orthogonal to $\vec {AB}$.

Time antisymmetry in antipodal directions is also found in a consistent quantum model\cite{Hooft:2016cpw,Hooft:2016itl,Hooft2018}   of  inbound and outbound particle states of  an eternal quantum black hole event horizon (Fig. \ref{blackhole}).  That particular model does not explicitly treat holographic directional correlations, as it only quantizes the radial part of the back-reaction of quantized particle states on the metric. On the other hand, directional quantization must exist in some form to be consistent with black hole entropy.  If the hypothesis of this paper about the inflationary horizon is correct,  the nonlocal holographic information in quantized states of black hole horizons should produce directional entanglement, correlations and fluctuations similar to those discussed above (e.g., Eq. \ref{perpvariance}).

A similar relation was used in ref.\cite{0264-9381-35-20-204001} to model  experimental cross spectra of interferometers. In that application, the  cross spectrum of two signals is imaginary Planck amplitude noise,  filtered on a scale $R$ determined by a set of mirrors used to project  directional states of propagating light  onto a data stream in the classical proper time of a single laboratory rest frame. If  spooky cosmological correlations are detected, it is likely that  Planck scale correlations could be measured in  suitably configured experiments\cite{Holo:Instrument,2013PhRvL.110u3601R,PhysRevA.92.053821,Pradyumna:2018xbx}.

In our extrapolation to  inflation,  classical cosmic time  is determined by  the time component ($\nu=0$) of the  classical timelike vector $u_\phi^\nu$ defined for each world line by the unperturbed inflationary metric, as discussed above (Eq. \ref{fourD}).
The antisymmetry in observable 3D spatial perturbations follows from general covariance, by projection into  3D comoving transform space (Eqs.  \ref{fourD} and \ref{triple}).  Our model is that the   fluctuations become ``frozen in time" when an emergent perturbation crosses the horizon, leaving an image on the frozen classical metric of antisymmetric Planck amplitude noise  filtered at $f\approx H$.

The argument just given refers to a classical laboratory time $t$, but a consistent  theory must define relations between concrete observables.
It is  useful to contrast  our  quantum-time measurement  with  Einstein's classical thought experiment in which  ``light clocks'' measure ticks of local proper time by bouncing light between mirrors at fixed separation in a rest frame.
That experiment shows how relativistic time dilation occurs:   two observers in  relative motion who compare clocks both see the other's clock ticking slowly,  because light has to travel farther to accommodate apparent position displacements in the moving frame. 
 Here, the output of a  light clock corresponds to a directionally oriented time operator, similar to  $\hat \delta_{i\pm}$ or $\Delta {\hat \tau}_i$; 
signals of  light clocks in different directions do not commute, and comparisons of three directions obey a spin-like algebra like that studied above, where the  operator $\hat T$  represents classical proper time.
The time series $\delta_{AB}(t)$,  $\delta_{BA}(t)$ can be operationally defined as  differences between clocks aligned orthogonally to the $AB$ spatial separation direction, near  $A$ and and $B$ world lines respectively.
Thus,  the exotic relative time  fluctuation is not always  a dilation, but can have either positive or negative sign,  averages to zero for two observers at rest, and is directionally antisymmetric in both space and time.

\end{document}